\documentclass[twocolumn,english,aps,prl,noshowpacs,superscriptaddress,amssymb,amsfonts]{revtex4-1}
\usepackage[T1]{fontenc}
\usepackage[latin9]{inputenc}
\usepackage{amsmath}
\usepackage{epstopdf}
\usepackage{graphicx}
\usepackage{amssymb}
\usepackage{color}
\usepackage{bm}
\usepackage{mathrsfs}
\usepackage[low-sup]{subdepth}
\usepackage{multirow}
\usepackage{amsfonts}
\usepackage[english]{babel}
\usepackage[T1]{fontenc}
\usepackage{times}
\usepackage[colorlinks,bookmarks=true,citecolor=blue,linkcolor=red,urlcolor=blue]{hyperref}
\usepackage[tight, FIGTOPCAP, hang, raggedright, nooneline]{subfigure}
\usepackage{verbatim}
\usepackage{dcolumn}
\usepackage{bbold}
\usepackage{babel}

\makeatletter
\newcommand{\be}{\begin{equation}}
\newcommand{\ee}{\end{equation}}
\newcommand{\bea}{\begin{eqnarray}}
\newcommand{\eea}{\end{eqnarray}}

\providecommand{\up}{\uparrow}
\providecommand{\dn}{\downarrow}

\newcommand{\ket}[1]{\left|#1\right>}
\newcommand{\bra}[1]{\left<#1\right|}
\newcommand{\braket}[1]{\left<#1\right>}
\newcommand{\para}[1]{\left(#1\right)}
\newcommand{\abs}[1]{\left|#1\right|}

\newcommand{\COMMENT}[1]{}

\def\beq{\begin{equation}}
\def\eeq{\end{equation}}
\def\barray{\begin{eqnarray}}
\def\earray{\end{eqnarray}}
\def\up{\uparrow}
\def\dn{\downarrow}

\begin{document}

\title{A unified theory of variational and quantum Monte Carlo methods and beyond}
\author{Mohammad-Sadegh Vaezi}
\affiliation{Pasargad Institute for Advanced Innovative Solutions, Tehran, Iran}
\affiliation{Department of Physics, Washington University, St. Louis, MO 63160, USA}
\author{Abolhassan Vaezi}
\email{vaezi@stanford.edu}
\affiliation{Department of Physics, Stanford University, Stanford, CA 94305, USA}
\affiliation{Stanford Center for Topological Quantum Physics, Stanford University, Stanford, California 94305-4045, USA}
\begin{abstract}
We present a unified theory of the variational Monte Carlo (VMC) and determinant quantum Monte Carlo (DQMC) methods using a novel density matrix formulation of VMC. We introduce an efficient algorithm for VMC to compute correlation functions and expectation values based on the auxiliary field Hirsch-Hubbard-Stratonovic transformation. We show that this new approach to VMC converges significantly faster than its traditional implementations. Furthermore, we generalize the Trotter-Suzuki decomposition to finite imaginary time steps $\tau \sim O(1)$ and develop a variational quantum Monte Carlo (VQMC) method accordingly, which is more accurate than VMC and can incorporate quantum fluctuations more efficiently. The two extreme limits of the VQMC method, namely infinitesimal and infinite imaginary time steps, correspond to the DQMC and VMC techniques, respectively. We demonstrate that our VQMC allows us to access lower temperatures in comparison with the conventional DQMC before the sign problem comes into play. We finally show that our VQMC can also enhance the accuracy of the projector Monte Carlo methods by providing better and less biased candidates for the trial wave functions, requiring shorter projection times for a given accuracy and alleviating the sign problem further. 
\end{abstract}

\maketitle

\section{I. Introduction}

The exact solution of quantum mechanical problems is in general highly nontrivial and in most cases intractable, mainly  due to the exponential growth of the Hilbert space dimension with the system size. This obvious fact entails for reliable though approximate solutions of quantum problems. In the past, several reliable approaches capable of solving non-integrable models with arbitrary accuracy have been developed. For example, it has been shown that the ground-state of local (quasi-)one-dimensional gapped Hamiltonians can be obtained through the density-matrix-renormalization group and related quantum entanglement based methods with a finite computational effort independent of the system size~\cite{White_DMRG_1992a,Verstraete_Review_2008a,Orus_Review_2014a,Schollwock_Review_2011a}. Furthermore, quantum Monte Carlo (QMC) methods can solve several non-integrable models in higher dimensions, e.g., the Hubbard model on bipartite lattices at half-filling~\cite{Scalapino_QMC_1981a,White_QMC_1989a,Assaad_PMC_2008a,Berg_QMC_2018a,Sandvik_SSE_1991a,Assaad_Nature_2010a,Berg_QMC_2012a,Lederer_QMC_2015,Vaezi_QMC_2017a}. Nevertheless, these popular methods are inapplicable in most cases. For example, DMRG fails to solve higher-dimensional problems~\cite{Verstraete_Review_2008a,Orus_Review_2014a,Schollwock_Review_2011a}, or QMC away from half-filling or on frustrated lattices is plagued by the notorious fermionic sign problem~\cite{Huang_Science_2017a,Zhang_QMC_1995a,Wu_SP_2005a,Li_SP_2015a,Wang_SP_2015a,Chandrasekharan_SP_2013a}. The average sign of fermion determinants  in these situations vanishes as $\exp\para{-V f/T}$, where $V$ denotes the size of system, $T$ temperature, and $f$ is a constant dependent on the interaction strength. Since the number of required Monte Carlo Samplings for a given accuracy scales as $1/\braket{{\rm sign}}^2$, the QMC approach becomes infeasible at low temperatures or for large systems for generic models~\cite{Troyer_SP_2005a,Troyer_SP_2014a,Chandrasekharan_SP_1999a,Umriger_SP_2007a}. In this paper, we try to ameliorate the sign problem by first reformulating the variational Monte Carlo (VMC) method~\cite{Gros_VMC_1989a,Paramekanti_VMC_2001a,Clark_VMC_2011a,Sandvik_VMC_TPS_2007a} (which does not capture quantum fluctuations efficiently)  using a density matrix approach. We then generalize our VMC formulation to incorporate quantum fluctuations as well. The resulting variational quantum Monte Carlo (VQMC) unifies the determinant QMC (DQMC), and VMC methods as the two extreme limits of a single computational approach. Finally, we demonstrate that our VQMC can yield highly accurate results when combined with the projector QMC (PQMC) method~\cite{Assaad_PMC_2008a,Zhang_QMC_1995a} through providing better approximate trial ground states.

In this paper, we first show that our density matrix formulation of VMC allows an auxiliary field implementation. Based on this observation we introduce a more efficient algorithm for VMC. The resulting implementation converges faster than its conventional counterparts in coordinate space. Furthermore, we show that measuring correlation functions and expectation values of various operators is more convenient within this new framework. We then discuss our VQMC approach, in which we introduce a generalized Trotter-Suzuki decomposition associated with finite imaginary time steps. The resulting decomposition can be interpreted as a renormalization group transformation along the imaginary time direction. We demonstrate that such coarse graining operations will renormalize various coupling constants of the model Hamiltonian. For example, using the Hubbard model to benchmark our VQMC approach, we show that the onsite Hubbard coupling $U$ flows to weaker values upon considering larger imaginary time steps. We show that our VQMC yields ground-state properties with a higher accuracy compared to VMC. Furthermore, we argue that VQMC captures low energy quantum fluctuations and despite suffering from the fermioinic sign problem (similar to DQMC), the onset of the sign problem emerges at lower temperatures. Having achieved a better ansatz for the ground-state density matrix through applying VQMC, we can feed it into the PQMC machinery as its trial/guiding state and achieve ground-state properties with high confidence using considerably shorter projection times. The latter observation can potentially circumvent the fermionic sign problem.

In this paper, we study the Hubbard model~\cite{Hubbard_1963a} on the square lattice defined as $H = H_K + H_U$, where its kinetic part: $H_k=-t_1\sum_{\braket{ij},\sigma}c_{i,\sigma}^\dag c_{j,\sigma} $ and the interaction part: $H_U = U\sum_{i,\sigma}n_{i,\uparrow}n_{i,\downarrow}$. In these expressions, $\braket{ij}$ denotes nearest neighbors, $c_{i,\sigma}$ electron annihilation operator on site $i$ with spin $\sigma$, and $n_{i,\sigma} = c_{i,\sigma}^\dag c_{i,\sigma}$ the corresponding electron number.

\section{II. A unified theory of variational, projector and quantum Monte Carlo methods}

In this section, we present a unified formulation of the variational and (determinant) quantum Monte Carlo methods. Based on the acquired knowledge we establish a novel technique dubbed as variational quantum Monte Carlo that paves the way between VMC and DQMC. We finally demonstrate that our VQMC-PQMC hybrid approach can outperform all existing QMC methods through providing more accurate initial density matrix (trial state) in the PQMC method. 

\subsection{A. Density matrix formulation of the VMC method}
In the conventional VMC approach, we try to come up with a guess for the ground-state wave function. Let us denote the true ground-state by $\ket{\Psi_G}$, and the trial one by $\ket{\Phi_T}$. The guessed/trial wave function $\ket{\phi_T}$ contains a number of variational parameters, which are tuned to minimize the expectation value of the model Hamiltonian with respect to $\ket{\Phi_T}$.  The variational wave function is usually defined through a non-interacting fermion wave function projected by a Gutzwiller-Jastrow operator, namely: $\ket{\Phi_T} = {Z}^{-1/2}P_G \ket{\Phi_0}$, where $Z$ is a normalization factor, and the (partial) Gutzwiller projection operator is defined as
\beq
P_G = \prod_{i} \para{1- \tilde{g} n_{i,\up}n_{i,\dn}} =  e^{- \frac{g}{2} \sum_{i}n_{i,\up}n_{i\dn}},\label{eq:GJ-1} 
\eeq
in which $g = -2\log(1-\tilde{g})$ serves as a variational parameter (which controls the onsite double occupancy). The non-interacting fermionic  wave function $\ket{\Phi_0}$ is the ground-state of a {\em variational} quadratic fermion Hamiltonian $H_{M} = -\sum_{ij} \psi_{i}^\dag {M_{ij}} \psi_j$, where $\psi_{i} = \para{c_{i,\up},c_{i,\dn}^\dag}^{\rm T}$ denotes the spinor fields (in Nambu space). Therefore, $H_{M}$ can in general accommodate pairing terms as well. The main task in VMC is to minimize the expectation value of the Hubbard Hamiltonian w.r.t. the variational  wave function, i.e., to minimize $E_{g,M} = \bra{\Phi_T}H\ket{\Phi_T}$ functional, from which  $g$ and elements of the variational hopping matrix $M$ can be obtained. In the conventional implementation of VMC, the variational  wave function is written in real space using Slater determinants and the expectation values are computed using Monte Carlo algorithm in which electron coordinates are sampled via  Metropolis-Hasting importance sampling~\cite{Metropolis_MC_1953a,Hasting_MC_1970a}. In this approach, even for $g = 0$, we need to sample thousands of configurations to achieve a fair estimate of the expectation value of various operators.

In this paper, we take a different route and instead of working with  wave function and Slater determinants, we work with density matrix operators. Our main observation is the following identity:
\begin{eqnarray}
&& \rho_{\rm VMC} =  \ket{\Phi_T}\bra{\Phi_T} = \frac{1}{Z} P_{G} \ket{\Phi_0}\bra{\Phi_0} P_{G} \cr
&&~~~=  \frac{1}{Z}  e^{-\frac{g}{2}\sum_{i}n_{i,\up}n_{i,\dn}} \para{\lim_{\beta \to \infty} e^{-\beta \psi^\dag {\bf M }\psi} }e^{-\frac{g}{2}\sum_{i} n_{i,\up}n_{i,\dn}},\label{eq:VMC-1} ~~~~
\end{eqnarray}
where we have used the fact that the density matrix associated with noninteracting electrons is unique and identical to the Boltzmann operator when $T = 1/\beta \to 0$. The above seemingly trivial identity has a profound implication. Indeed, it suggests the following approximate generalization of the Trotter-Suzuki decomposition for gigantic time steps:
\beq
{\rm VMC:}~~~~~~ e^{-\beta \para{H_{K} + H_{U}}} \approx e^{-\beta H_{U_{\rm eff}}/2}  e^{-\beta H_{M}} e^{-\beta H_{U_{\rm eff}}/2},\label{eq:Trotter-1} 
\eeq
where $U_{\rm eff} = g/\beta$ is the renormalized onsite Hubbard coupling. Although the above relation is clearly an approximation and very far from being accurate, it is the basis of the VMC method and all of its successes as well as limitations. We are now prepared to utilize Monte Carlo methods to evaluate the approximate VMC density matrix $\rho_{\rm VMC}$. To this end, similar to DQMC, we first use the discrete Hubbard-Stratonovic transformation introduced by Hirsch~\cite{Hirsch_HS_1983a} using the following identity:
\beq
e^{-\frac{1}{2}g n_{i,\up}n_{i,\dn}} = \frac{1}{2}e^{-\frac{g}{4}\para{n_{i,\up}+n_{i,\dn}}} \sum_{s_{i}=\pm 1}e^{\lambda s_{i}\para{n_{i,\up}-n_{i,\dn}}},\label{eq:HOS-1} 
\eeq
where $\cosh\para{\lambda} = \exp\para{g/4}$. The above Hubbard-Stratonovic transformation can be employed to rewrite the VMC density matrix in terms of an ensemble of noninteracting (i.e., quadratic) density matrices over the right and left Hubbard-Stratonovic auxiliary binary fields as follows:
\begin{eqnarray}
\quad \rho_{\rm VMC} &&\para{\beta}  ~~\propto \sum_{\left\{s^{L}_{i}\right\}} \sum_{\left\{s^{R}_{j}\right\}} \exp(-\lambda  \sum_{i,\sigma} \sigma s^{L}_{i} n_{i,\sigma})   \cr
&& \times \exp\para{-\beta \psi^\dag H_{M} \psi} \exp(-\lambda \sum_{j,\sigma}\sigma  s^{R}_{j}n_{j,\sigma}).\label{eq:VMC-2} 
\end{eqnarray}

Now, we can employ the Metropolis-Hasting algorithm to sample important configurations of the Hubbard-Stratonovic fields. In our calculations we consider large values of $\beta$ and extrapolate to $\beta \to \infty$. Computing expectation values is more convenient using the above algorithm compared to the conventional VMC implementations. For example, in a single step, we can easily obtain any expectation value for $g=0$ while the conventional implementation of the VMC method requires thousands of sweeps for the same parameters. On the contrary, our density matrix based method converges faster and expectation values can be obtained with high accuracy by sweeping over Hubbard-Stratonovic fields only a few thousand times even at the large $U$ limit.  
We would like to stress that the above scheme is spin-problem free due to the fact that the renormalized Hubbard coupling, i.e., $U_{\rm eff} = g/\beta$, is negligible (since $g/U\sim O(1)$) and it is known that such weak couplings do not lead to sign problem (see Figs. \ref{fig:Fig_4} and \ref{fig:Fig_7} ).

\subsection{B. Variational Quantum Monte Carlo}

As we argued above, the VMC method can be viewed as a generalized Trotter-Suzuki decomposition for very large imaginary time steps (see Eq. \ref{eq:Trotter-1}). In this section, we generalize VMC and find a smooth path connecting VMC to DQMC. Our main tool to achieve this goal is the following generalized Trotter-Suzuki decomposition for finite imaginary time steps $\tau \equiv \beta/N$:
\beq
e^{-\tau \para{H_K+H_U}}  \approx \rho^{\rm eff}_{\tau} \equiv e^{-\frac{\tau}{2} H_{s}} e^{-\frac{\tau}{2} H_{c}} e^{-\tau H_{M}} e^{-\frac{\tau}{2} H_{c}}   e^{-\frac{\tau}{2} H_{s}}, \label{eq:Trotter-2} 
\eeq
where $H_{M} = \psi^\dag{\bf  M} \psi$ denotes our variational quadratic Hamiltonian, and $H_{c} = 1/2\sum_{r}V_{r}n_{i}n_{i+r}$ and $H_{s} = \sum_{r}J_{r}{\bf s}_{i}.{\bf s}_{i+r}$ are two body density-density and spin-spin interactions with variational couplings $V_r$ and $J_r$, respectively.
The above transformation must recover the original Trotter-Suzuki decomposition for infinitesimal time steps which requires $\tau \to 0 :~H_{M} \to H_{K}, ~~V_{0} \to U$ and all other components of $V$ and $J$ to vanish. Instead, for finite $\tau \sim O(1)$, we consider longer range two body interactions to improve the accuracy. Indeed, the long range density-density interaction considered above is a second quantized representation of the Jastrow factors known in the studies of the Bose gas. In this work, we truncate such long range terms to keep the discussion and calculations simple. The above generalized Trotter-Suzuki decomposition can be interpreted as a renormalization group transformation along the imaginary time direction where we coarse grain infinitesimal time steps all the way to the desired imaginary time scale $\tau \sim O(1)$. The immediate consequence of the above generalized Trotter-Suzuki decomposition is the following trivial identity:
\beq
e^{-\beta H} = \para{e^{-\tau H}}^N \approx \rho_{\rm VQMC} \equiv  \prod_{i = 1}^{N} \rho^{\rm eff}_{\tau},~~~\tau = \beta /N.\label{eq:VQMC-1} 
\eeq
Now, similar to the DQMC algorithm, we can apply the discrete Hubbard-Stratonovic transformation (see Eq. \ref{eq:HOS-1}) for each $\rho_{\tau}$ component, and rewrite the total density matrix as a path integral over the Hubbard-Stratonovic auxiliary fields. There are several methods to find the variational parameters considered in Eq. \ref{eq:Trotter-2}. The most straightforward one is to minimize the expectation value of the Hubbard Hamiltonian for large $\beta$ (at which entropic terms are negligible). 

\begin{figure}
\subfigure[]{\includegraphics[height = 4.8cm, width=6.5cm]{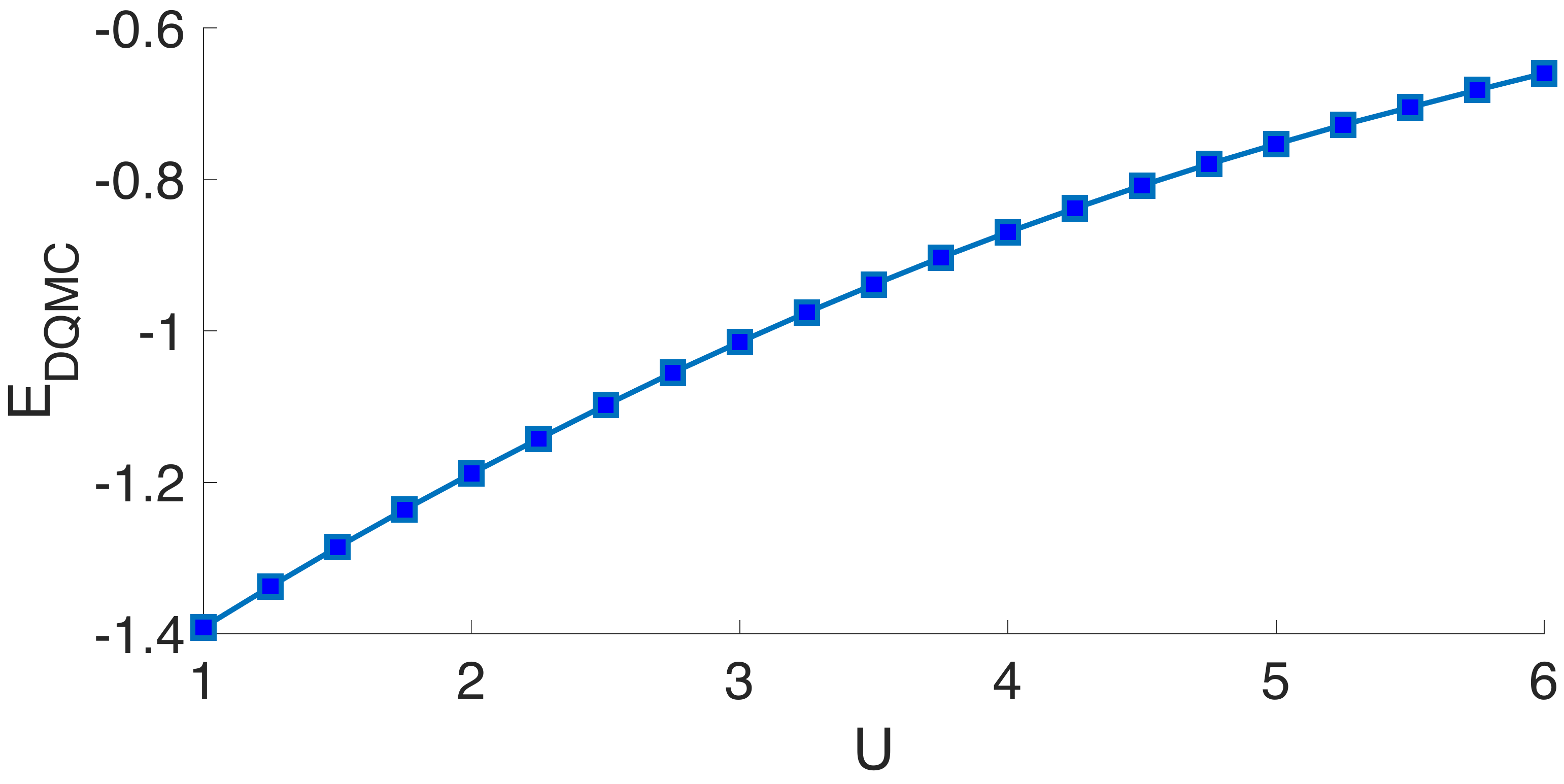}}
\subfigure[]{\includegraphics[width=6.5cm]{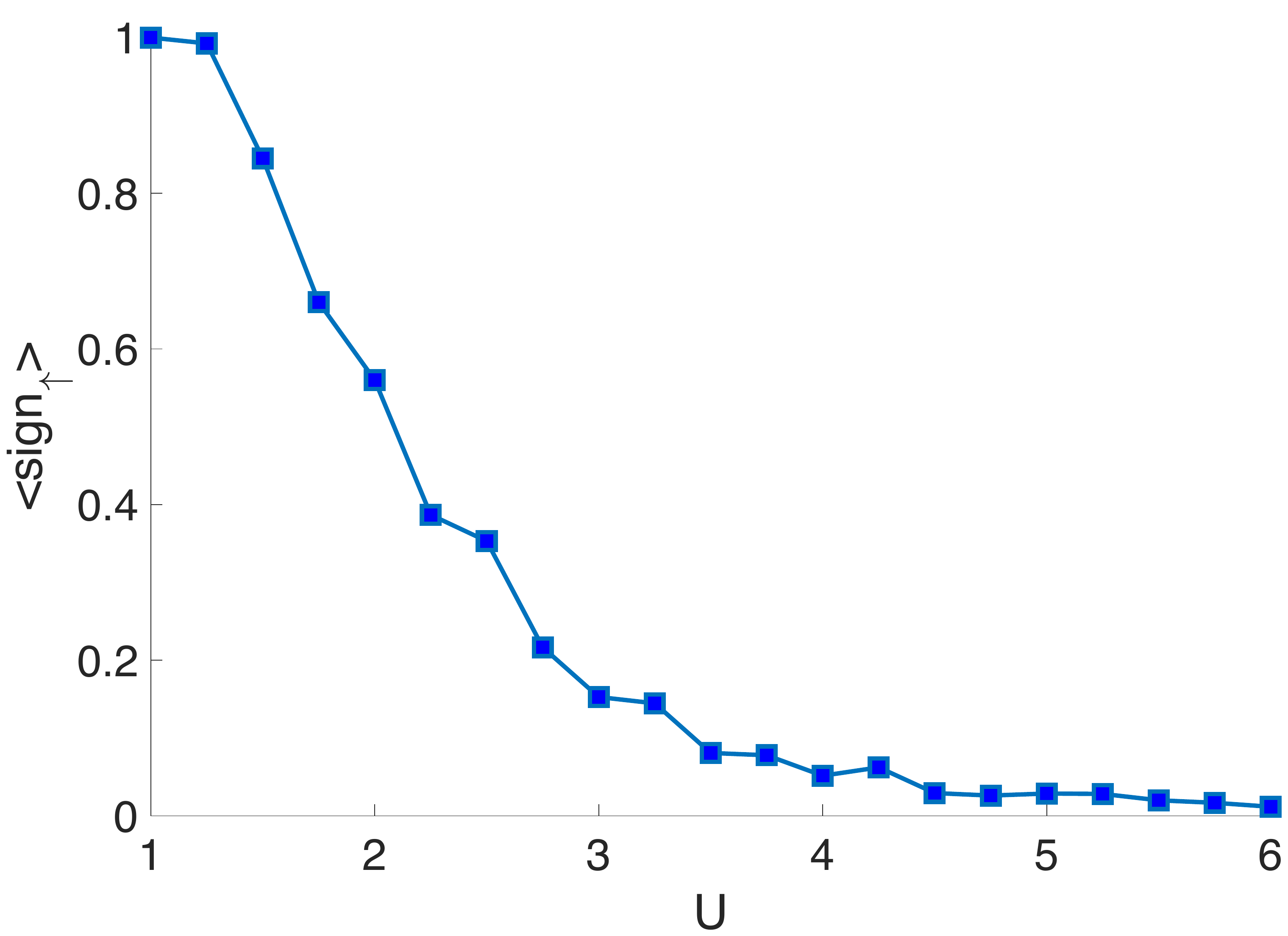}}
\caption{DQMC results for the half-filled Hubbard model. (a) Ground-state energy per site obtained by DQMC at $\beta = 20$ for a $16 \times 4$ square lattice as a function of onsite Hubbard interaction strength $U$. These values provide the exact energies for the ground-state (up to $0.0005$ statistical error bar) due to the absence of (overall) sign problem. (b) The average sign of spin up determinants for the same parameters. Although the overall sign problem is absent at half filling, the average sign of spin up fermion determinants is highly fluctuating especially at strong couplings.} \label{fig:Fig_1}
\end{figure}

There are four reasons to advocate for the above VQMC method: (i) It unifies two important tools in computational physics namely VMC and DQMC and identifies them as two extreme points of a unique method where DQMC corresponds to $\tau \to 0$, and VMC corresponds to $\tau = \beta \gg 1$. (ii) It is more accurate than VMC when $N > 1$. This is due to the fact that we have kept quantum fluctuations by partitioning the imaginary time direction into $N $ slices. (iii) Consequently, VQMC allows measuring unequal (imaginary) time correlation functions and by using analytic continuation~\cite{Jarrel_analytic_QMC_1996a} we can achieve useful insights for the real time dynamic of our system. More specifically, in this scheme we can define Matsubara frequencies as $\omega_n = \frac{2\pi}{\beta}n$, where $n = 1,\cdots N$, thus we have kept and considered quantum fluctuation up to $\omega_{\rm max} = \frac{2\pi}{\tau}$ energy scale. (iv) Similar to DQMC, considering $N \gg 1$ time slices leads to fermionic sign problem, even for $\tau \sim O(1)$. However, the onset of $\beta$ above which the sign problem becomes significant is substantially larger than that of the conventional DQMC (see Figs. \ref{fig:Fig_1} and \ref{fig:Fig_7}). This allows us to study the ground-state properties before the sign problem comes into play.

We now argue that although VQMC suffers from the fermionic sign problem in general, which is a consequence of taking quantum fluctuations into account, it does not show up down to extremely low temperatures. This is mainly due to the fact that the renormalized onsite Hubbard interaction is notably weaker than its bare value for $\tau \sim O(1)$ as shown in Figs. \ref{fig:Fig_2_3} and \ref{fig:Fig_4}. This pivotal observation along with its higher accuracy and potentials, suggests the VQMC method as a competitive tool for studying strongly interacting fermion systems, especially when combined with the projector quantum Monte Carlo as will be explained in detail later in this section.

To demonstrate the absence of the sign problem down to low enough temperatures, we consider the half filled Hubbard model in this paper. Although the Hubbard model at half filling and on the bipartite lattices does not suffer from the notorious fermionic sign problem, the average sign of spin up (or down) determinants is not positive definite and fluctuates strongly at low temperatures and strong interactions. The absence of the overall sign problem just implies that the sign of spin up determinant equals that of the spin down ones for each realization of the Hubbard-Stratonovic field configuration. Indeed, the emergence of the sign problem and it severity can be readily studied and predicted by studying the average sign of spin up/down determinants at half filling. Thus, in the following we consider the average sign of spin up/down determinants as a metric for measuring the severity of the (product) sign problem in the general case and away from half filling.

\subsection{C. Beyond VQMC: variational projector quantum Monte Carlo}  We now demonstrate that our VQMC can yield highly accurate results when combined with the projector quantum Monte Carlo (PQMC) method while mitigating the sign problem as a corollary. The main idea of the PQMC is the observation that we can achieve the ground state(s) via $\ket{\Psi_G} = \lim_{\beta_1 \to \infty}Z^{-1/2}e^{-\beta_1 H/2}\ket{\Phi_T}$ relation for any trial wave function $\ket{\Phi_T}$ that has nonzero overlap with the true ground state wave function i.e., ${\braket{\Psi_G|\Phi_T}} \neq 0$. Practically, this identity still holds for finite but large enough $\beta_1$. However, the convergence speed, i.e., the threshold for $\beta_1$ beyond which the identity holds up to a given accuracy, depends on $\abs{\braket{\Psi_G|\Phi_T}}^2$ overlap which measures the distance between the trial and exact ground states. The main advantage of the PQMC method is that the sign problem becomes less severe as we discuss below. It is mostly a consequence of starting with a good trial wave function which can impose certain constraints on the Hubbard-Stratonovic fields (see section IV). 

\begin{figure}
\includegraphics[height=5.9cm]{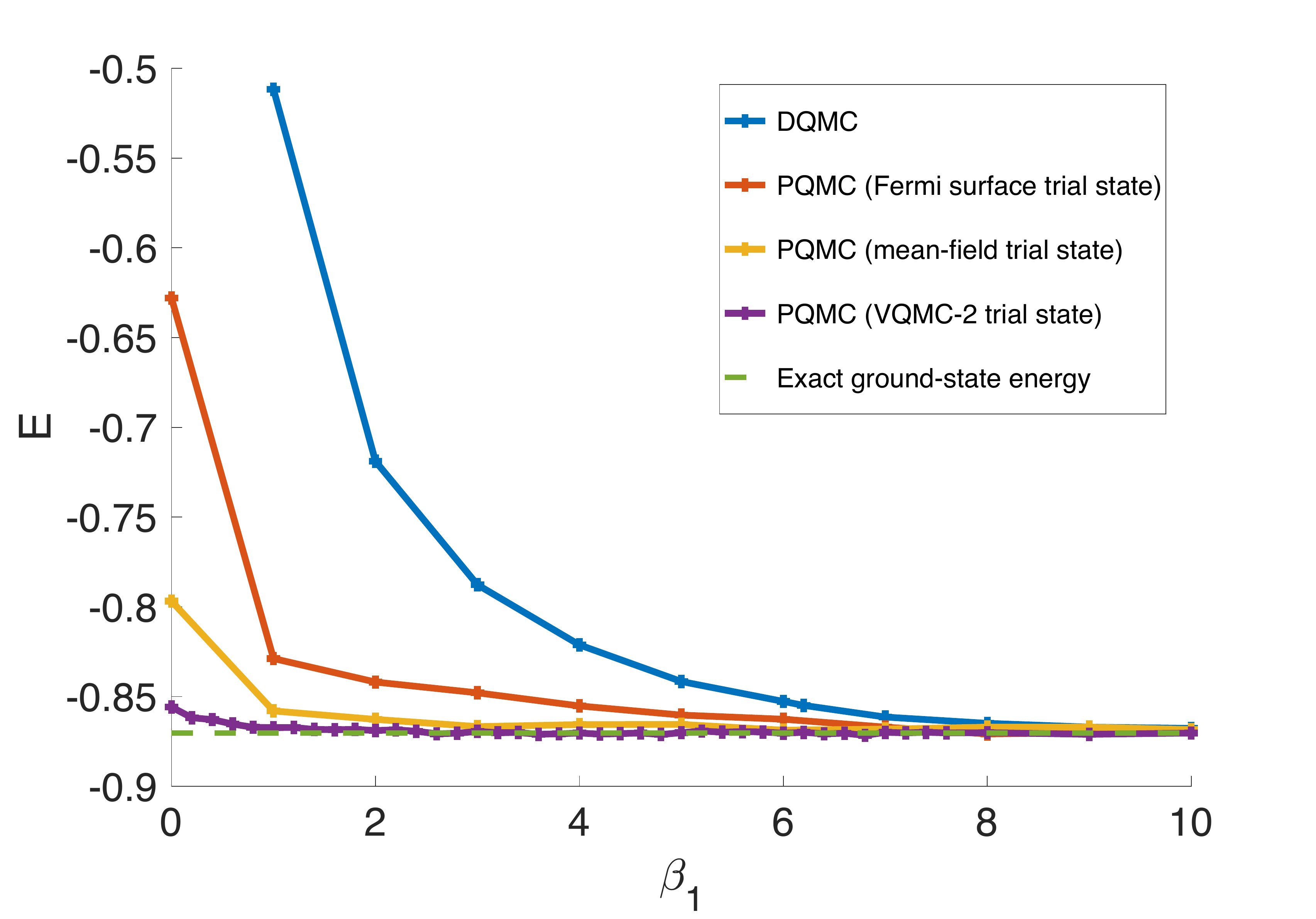}~~~
\caption{Ground-state energy estimated by PQMC with several different choices for the trial state. These results correspond to $U=4$ and $16\times4$ rectangular system. Note that PQMC converges faster than DQMC (which can be imagined as a PQMC with identity trial density matrix i.e. infinite temperature limit) even by starting from Fermi surface. Using VQMC with two parameters, $\tau = 1/2$, and $\beta_2 = 20$ as the trial density matrix yields an accurate energy estimation even for a projection time as short as $\beta_1/2$ = 1/2.} \label{fig:Fig_5}
\end{figure}

Similar to our derivation of Eq. $\ref{eq:VMC-1}$, the density matrix in PQMC can also be represented as: 
\begin{eqnarray}
\rho_{\rm PQMC} \equiv  ~&&\ket{\Psi_G}\bra{\Psi_G} \propto e^{-\beta_1 H/2}\ket{\Phi_T}\bra{\Phi_T}e^{-\beta_1 H/2} \propto\cr
&&\cr
&&   e^{-\beta_1 H/2}\rho_{T}\para{\beta_2} e^{-\beta_1 H/2},~~~\label{eq:PQMC-1} 
\end{eqnarray}
where ideally $\beta_{1,2} \to \infty$ though in practice we consider finite $\beta_{1,2}$. Furthermore, we can utilize the original Trotter-Suzuki decomposition to write the outer components as: $e^{-\beta_1 H/2} = \para{e^{-\tau_1 H}}^N_1$, and $e^{-\tau_1 H} \approx e^{-\tau_1 H_U/2} e^{-\tau_1 H_K} e^{-\tau_1 H_U/2}$ for $\tau_1 \ll 1/\sqrt{Ut_1}$. 
This decomposition enables us to employ the Hubbard-Stratonovic transformation and borrow other techniques from DQMC to evaluate various correlation functions (see section IV).
It is a common practice to first obtain a good trial wave-function $\ket{\Phi_{T}}$ using the mean-field or Hartree-Fock approximations and feed it in the above expression. Alternatively, in this paper we suggest using our VMC or VQMC methods to obtain a better trial density matrix $\rho_T\para{\beta_2}$ at temperature $T_2=1/\beta_2 \ll 1$. Using these alternative approaches, and in particular when $\tau_2 \sim O(1)$, leads to a shorter threshold for the projection time (i.e., $\beta_1/2$) due to the intrinsic higher accuracy of our VMC/VQMC than the mean-filed or Hartree-Fock approximations (see Fig. \ref{fig:Fig_5}). 
More importantly, in this framework we have the choice to use the ergodic form of VQMC where we impose all symmetries of the model Hamiltonian on $H_M$ in Eq. \ref{eq:Trotter-2}. Alternatively, we could allow symmetry breaking terms in $H_M$ but instead consider an ensemble over all possible degenerate ground-states.

Intuitively, we can estimate the threshold for $\beta_1$ such that $ 0 \leq E - E_{\rm G} < \epsilon$, where $E$ and $E_{\rm G}$ are the estimated and exact ground-state energies, respectively as follows. Let us assume the average energy of our trial density matrix $\rho_{\rm T}\para{\beta_2}$ is $E_2$. Furthermore, let us assume that we could reach $E_2$ in the usual finite temperature DQMC by considering an effective inverse temperature $\beta^{\ast}$. This suggests the following approximation: $\rho_{\rm T}\para{\beta} \approx e^{-\beta^{\ast}H}$, which in turn implies: $\rho_{\rm PQMC}\para{\beta_1,\rho_{\rm T}\para{\beta_2}} =  e^{-\beta_1 H/2}\rho_{T}\para{\beta_2} e^{-\beta_1 H/2} \approx  e^{-\para{ \beta_1 + \beta^{\ast}}H} = \rho_{\rm DQMC}\para{\beta_1+\beta^{\ast}}$. Thus, the density matrix of the PQMC method is related to that of the DQMC at a higher effective inverse temperature $\beta_1 + \beta^{\ast}$. Now, suppose that in the usual finite temperature DQMC method we could reach the desired accuracy $\epsilon$ in ground-state energy estimation by considering $\beta \geq \beta^{th}\para{\epsilon,\rm DQMC}$. Consequently, the threshold for $\beta_1$ in PQMC is $\beta^{ th}\para{\epsilon,\rm PQMC} = \beta^{\rm th}\para{\epsilon, \rm DQMC} - \beta^{\ast}$. Thus, $\beta^{ th}\para{\epsilon,\rm PQMC} < \beta^{ th}\para{\epsilon,\rm DQMC}$.  This observation suggests that the average sign of fermion determinants in PQMC is comparable to that of the DQMC method at a reduced inverse temperature, shifted down by $\beta^{\ast}$, and therefore the sign problem is less significant in PQMC and has an exponentially larger average sign assuming $\rho_{\rm T}\para{\beta_2}$ does not aggravate  the sign problem. Indeed, symmetry breaking choices of $\rho_{\rm T}\para{\beta_2}$ can elevate the average sign since they suppress quantum fluctuation/tunneling between degenerate ground-states.

Considering the above argument,  our main message in this paper is that  {\it using VQMC with $\tau_2 \sim O(1)$ leads to the highest possible value for $\beta^{\ast}$ without causing sign problem up to large enough values of $\beta_2$.  As a result, it minimizes $\beta^{ th}\para{\epsilon,\rm PQMC}$ and maximizes the average sign better than all other existing algorithms}. Therefore, it allows us to unravel previously unexplored ground-state properties of models that suffer from the sign problem.

\section{III. Implementation of our unified QMC algorithm}

In this section, we comment on the implementation of our single recipe which unifies several Monte Carlo approaches, namely VMC, its generalization VQMC, DQMC, and PQMC methods. Our starting point is the following identity which was derived before:  $\ket{\Psi_G}\bra{\Psi_G} = \lim_{\beta_{1,2} \to \infty} \rho$, where $\rho$ is defined as:
\beq
\rho \equiv e^{-\para{\beta_1+\beta_2}H} = e^{-\beta_1 H/2} e^{-\beta_2 H} e^{-\beta_1 H/2}.
\eeq
In the following, we consider $\beta_1 \ll \beta_2$ limit. We now break $\beta_1$ ($\beta_2$) into $N_1$ ($N_2$) time steps and arrive at the following trivial expression for $\rho$:

\beq
\rho = \para{e^{-\tau_1 H}}^{N_1/2} \para{e^{-\tau_2 H}}^{N_2} \para{e^{-\tau_1 H}}^{N_1/2}.\label{eq:rho-PQMC-1}
\eeq
Now, we consider $\tau_1$ ($\tau_2$) imaginary time steps to be infinitesimal (finite). Therefore, we can employ the conventional Trotter-Suzuki decomposition for $e^{-\tau_1 H}$ as follows:
\beq
\tau_1 \ll 1:~ e^{-\tau_1 H} \approx e^{-\tau_1 H_U/2}e^{-\tau_1 H_K}e^{-\tau_1 H_U/2}+O\para{\tau_1^3},\label{eq:Trotter-3}
\eeq
while for $e^{-\tau_2 H}$ we have to use the generalized Trotter-Suzuki decompositions introduced in Eq. \ref{eq:Trotter-2}. Again, for simplicity, we assume that $V_{r>0} = J_{r>0} = 0$. At the end of this section, we comment on how to include them in our algorithm. The above expression for $\rho$ corresponds to different QMC approaches depending on the values of $\para{\beta_1,\beta_2}$, and $\para{\tau_1,\tau_2}$. More specifically, {\it (a) DQMC corresponds to $\beta_1 > 0, \beta_2 = 0$ (note that $\tau_1 \to 0$, (b) VMC corresponds to $\beta_1 = 0,\beta_2>0$, and $\tau_2 = \beta_2$  (i.e., $N_2 = 1$). (c) VQMC corresponds to $\beta_1 = 0,\beta_2 >0$, and $N_2 > 1$, and (d) PQMC corresponds to $\beta_1 > 0, \beta_2 > 0$}. 

Having achieved the above (generalized) Trotter-Suzuki decompositions, we can employ the Hubbard-Stratonovic transformation (see Eq. \ref{eq:HOS-1}) after which the total density matrix $\rho$ can be written as an ensemble over non-interacting fermion density matrices subject to space-time dependent auxiliary binary fields. Accordingly, we obtain the following expression for $Z = {\rm Tr} \rho$

\begin{figure}
\subfigure[]{\includegraphics[height=3.05cm]{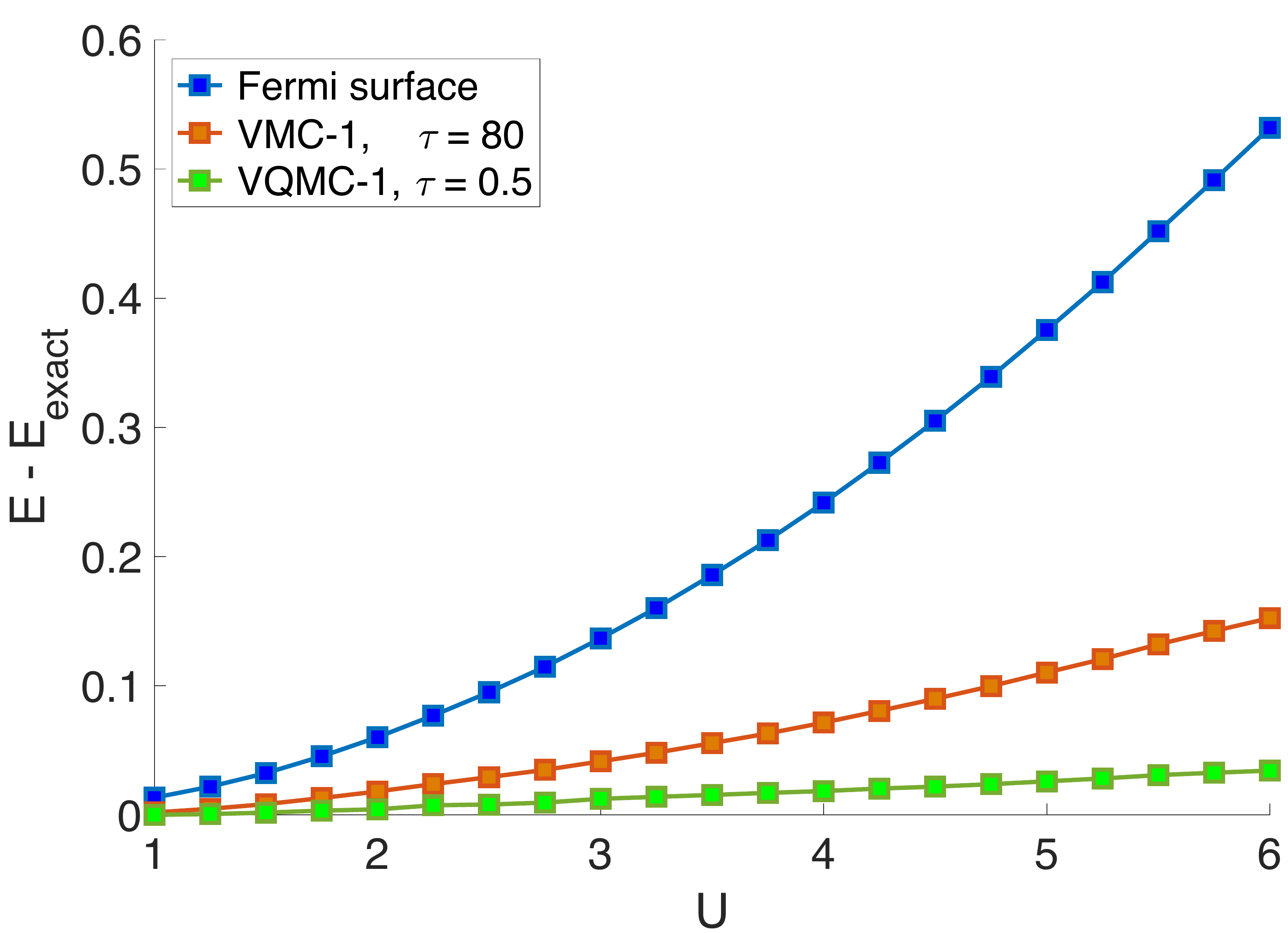}}
\subfigure[]{\includegraphics[height=3.05cm]{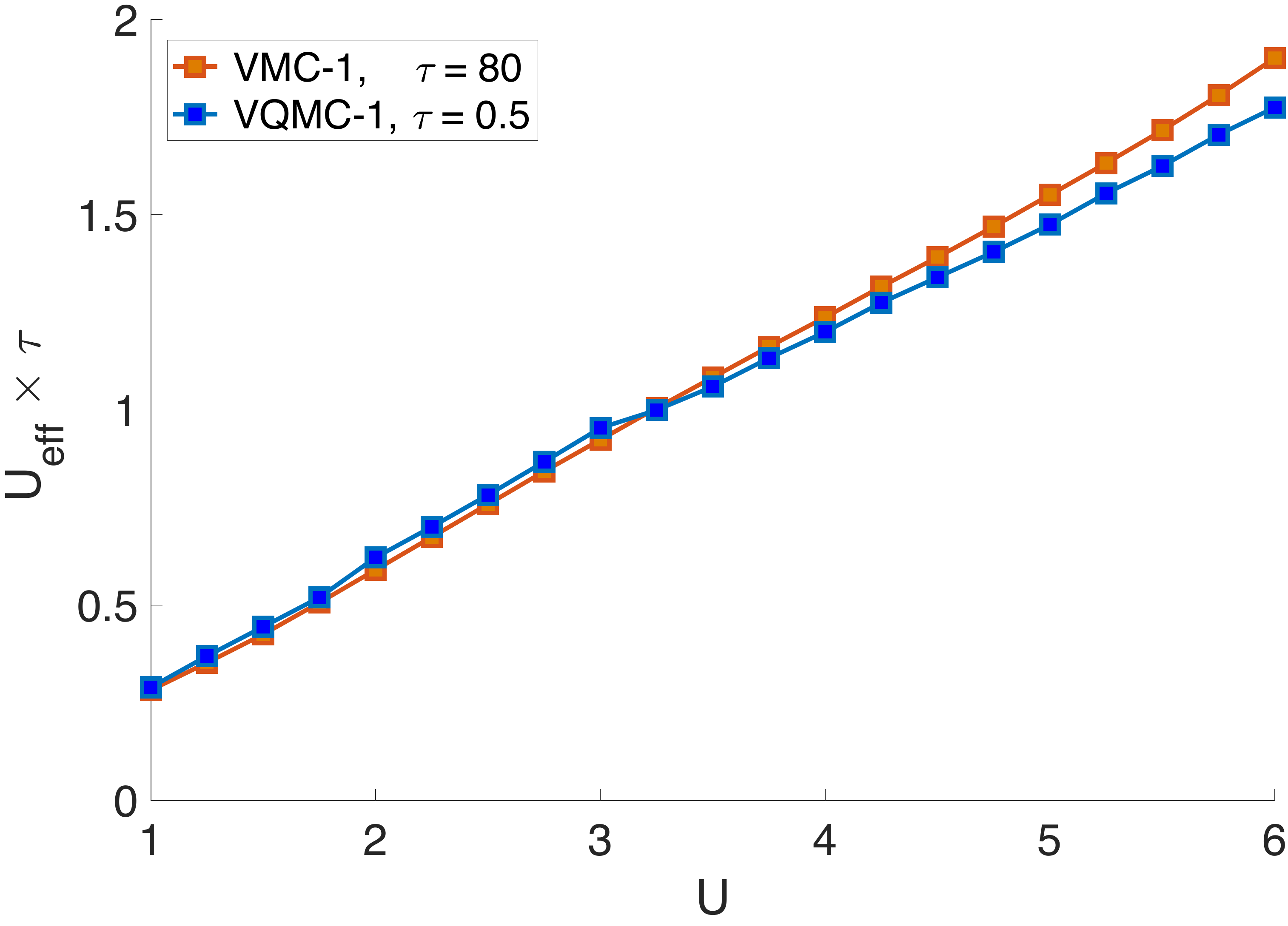}}
\subfigure[]{\includegraphics[height=3.05cm]{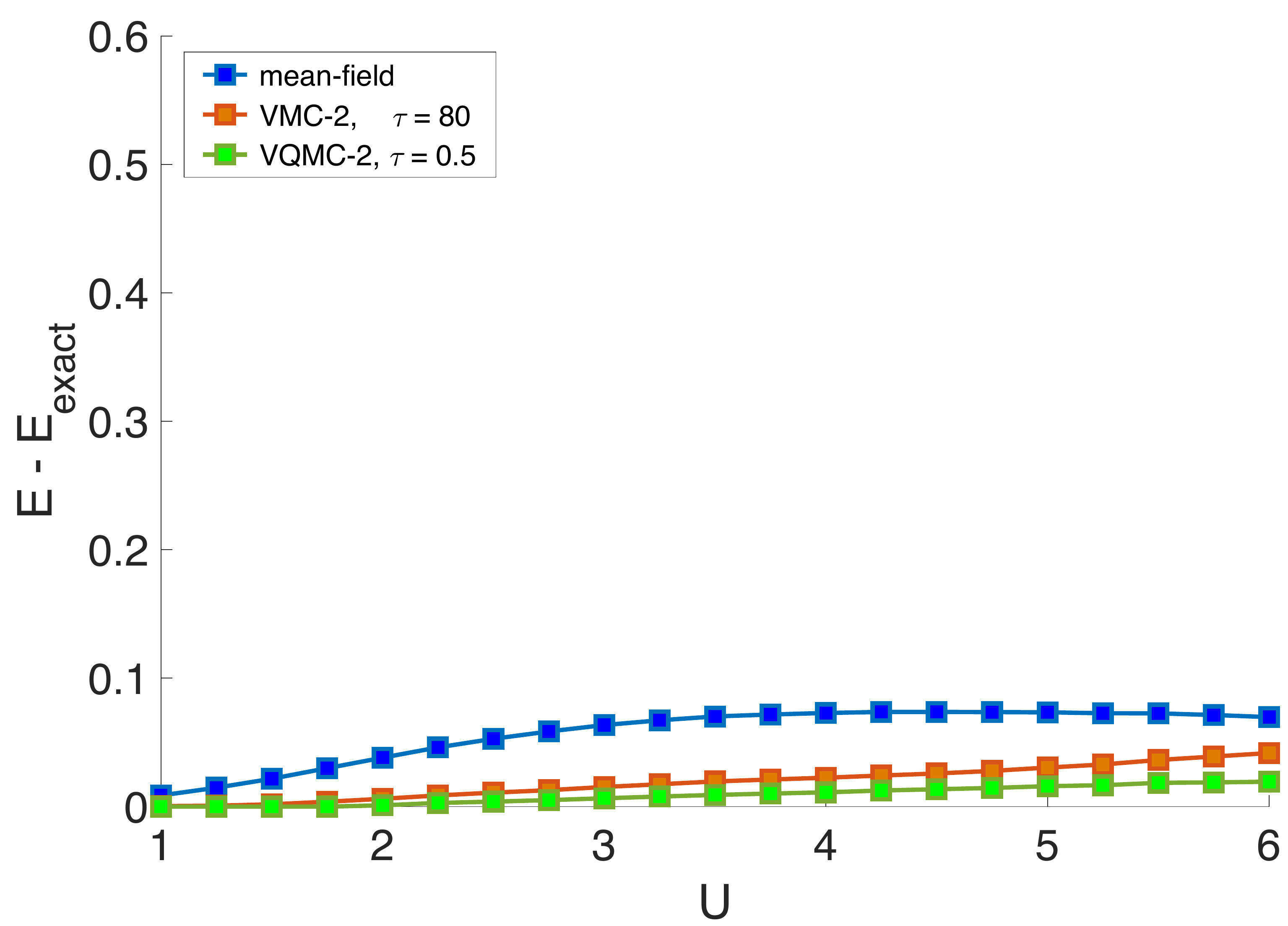}}
\subfigure[]{\includegraphics[height=3.05cm]{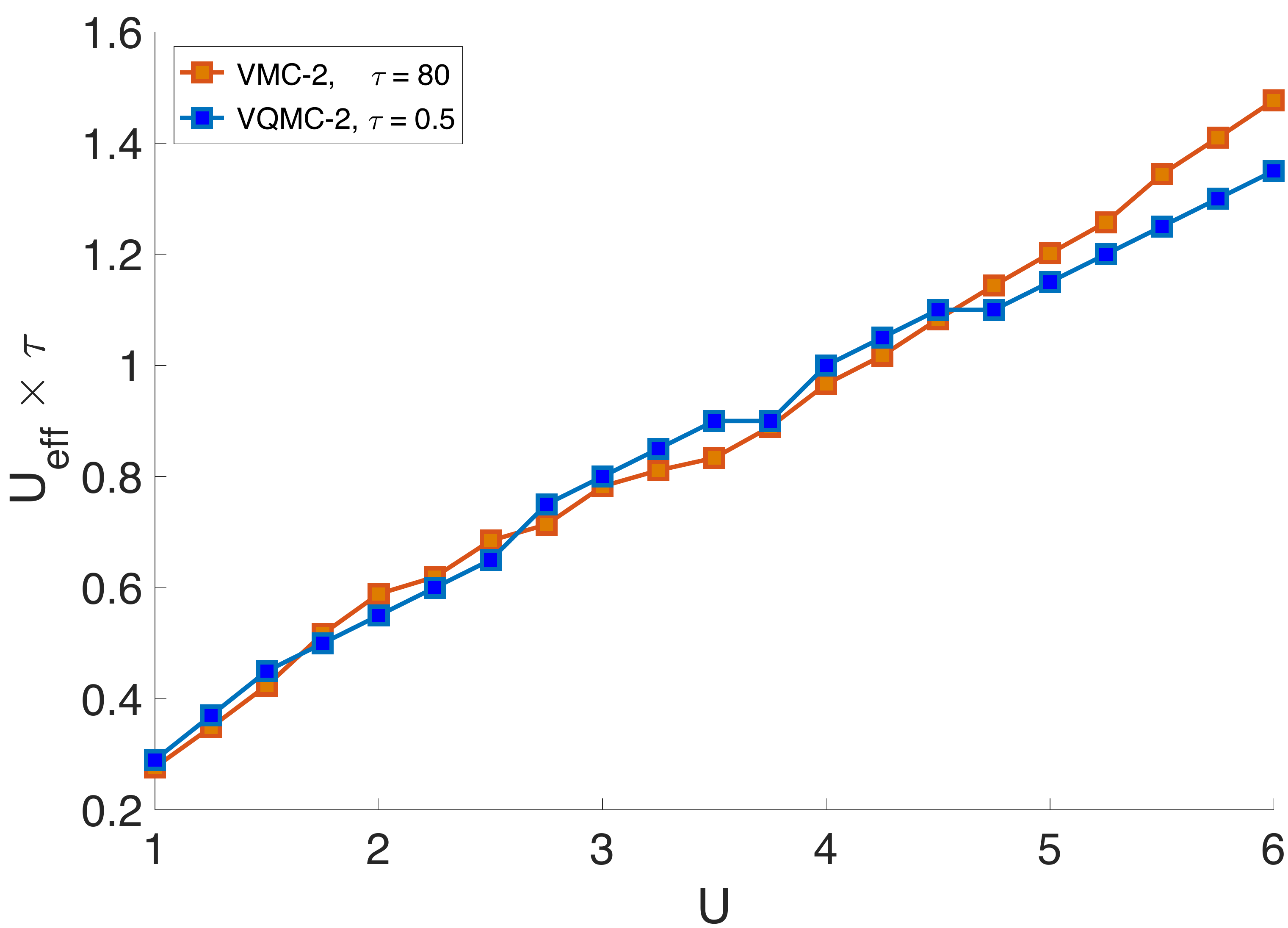}}
\caption{Results of one and two-parameter families of VMC and VQMC methods. (a) Gourd-state energy estimation of Free fermions (with square Fermi surface) as well as VMC and VQMC ansatz with one variational parameter $U_{\rm eff}$ (relative to the exact ground-state energy shown in Fg. 1). Here, we considered $\beta_2 = 80$ (40) for VMC (VQMC). (b) The corresponding optimal values of $U_{\rm eff}$. (c) and (d) plot similar quantities but now by optimizing two variational parameters: $U_{\rm eff}$ and $m$. These figures show that VMC provides an improvement over the noninteracting mean-field approximation, and VQMC improves VMC results further via taking quantum fluctuations into account. Also, considering more variational parameters yields more accurate results.} \label{fig:Fig_2_3}
\end{figure}

\begin{widetext}
\begin{eqnarray}
&&Z = \sum_{\{s^{1,L}\}} \sum_{\{s^{1,R}\}}\sum_{\{s^{2,L}\}}\sum_{\{s^{2,R}\}}  Z_{\uparrow}\para{\{s^{1,L}\},\{s^{1,R}\},\{s^{2,L}\},\{s^{2,R}\}}   Z_{\downarrow}\para{\{s^{1,L}\},\{s^{1,R}\},\{s^{2,L}\},\{s^{2,R}\}},
\end{eqnarray}
in which
\begin{eqnarray}
&& Z_{\sigma}\para{\{s^{1,L}\},\{s^{1,R}\},\{s^{2,L}\},\{s^{2,R}\}} = {\rm Tr}
\left[\hat{F}\para{{\bf s}^{1,L}_{N_1},{\bf s}^{1,R}_{N_1},\lambda_1,\sigma}\cdots\hat{F}\para{{\bf s}^{1,L}_{N_1/2+1},{\bf s}^{1,R}_{N_1/2+1},\lambda_1,\sigma}\times\right.\cr
&&~~~~~~~~~~~~~~~~~\left.  \hat{G}\para{{\bf s}^{2,L}_{N_2},{\bf s}^{2,R}_{N_2},\lambda_2,\sigma}\cdots \hat{G}\para{{\bf s}^{2,L}_{1},{\bf s}^{2,R}_{1},\lambda_2,\sigma}\hat{F}\para{{\bf s}^{1,L}_{N_1/2},{\bf s}^{1,R}_{N_1/2},\lambda_1,\sigma}\cdots\hat{F}\para{{\bf s}^{1,L}_{1},{\bf s}^{1,R}_{1},\lambda_1,\sigma}\right],
\end{eqnarray}
where $\hat{F}$ and $\hat{G}$ operators are defined as follows:
\begin{eqnarray}
&&\hat{F}\para{{\bf s}^{1,L}_{l},{\bf s}^{1,R}_{l},\lambda_1,\sigma} = e^{-\lambda_1\tau_1 s^{1,L}_{i,l}\sigma n_{i,\sigma}} e^{-H_K\tau_1} e^{-\lambda_1\tau_1 s^{1,R}_{i,l}\sigma n_{i,\sigma}},\quad \cosh\para{\lambda_1} = e^{U\tau_1/4}.\\
&&\hat{G}\para{{\bf s}^{2,L}_{l},{\bf s}^{2,R}_{l},\lambda_2,\sigma} = e^{-\lambda_2\tau_2 s^{2,L}_{i,l}\sigma n_{i,\sigma}} e^{-H_M\tau_2} e^{-\lambda_2\tau_2 s^{2,R}_{i,l}\sigma n_{i,\sigma}} ,\quad \cosh\para{\lambda_2} = e^{U_{\rm eff}\tau_2/4}.
\end{eqnarray}
Using standard relations for the product of fermion Gaussian forms, individual contributions to the partition function can be evaluated as:
\begin{eqnarray}
Z_{\sigma}\para{\{s^{1,L}\},\{s^{1,R}\},\{s^{2,L}\},\{s^{2,R}\}} =  \det\para{\mathbb{1}+B^{\sigma}_{N_1}\cdots B^{\sigma}_{N_1/2+1}A^{\sigma}_{N_2}\cdots A^{\sigma}_{1}B^{\sigma}_{N_1/2}\cdots B^{\sigma}_{1}},\label{eq:Z-1}
\end{eqnarray}
\end{widetext}

\begin{figure}
\includegraphics[height=5.5cm]{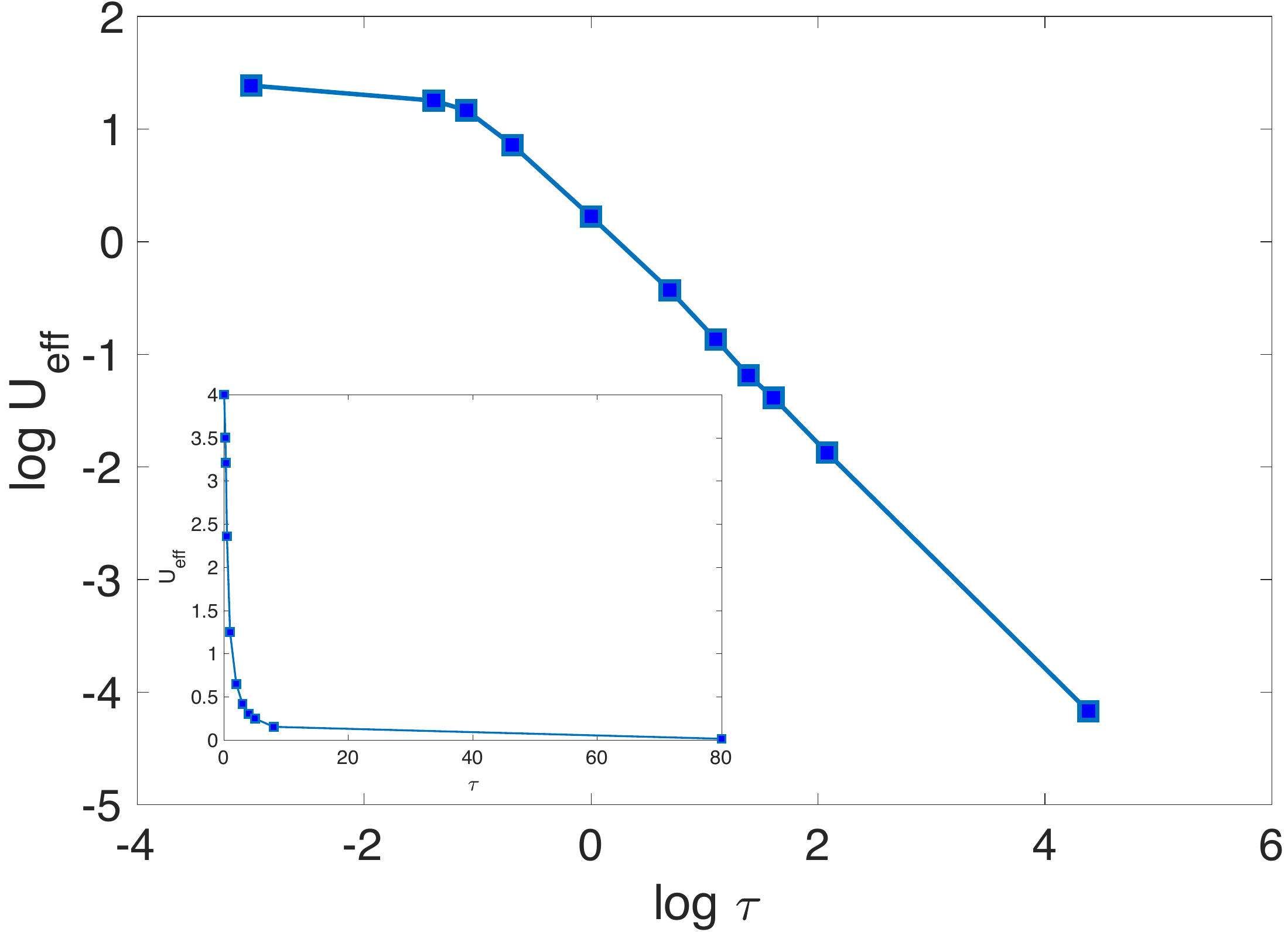}~~~
\caption{Renormalized onsite Hubbard interaction as a function of imaginary time steps (by optimizing one variational parameter only) for $U=4$. As this figure shows, $U_{\rm eff}\tau$ approaches a constant for $\tau \geq 1/2$.} \label{fig:Fig_4}
\end{figure}

where imaginary time dependent $A^{\sigma}_{l}$, and $B^{\sigma}_{l}$ matrices are defined as:
\begin{eqnarray}
&&A^{\sigma}_{l} = e^{-\sigma\lambda_1\tau_1 {\rm diag}\para{s^{1,L}_{i,l}}}   e^{-\tau_1 K^{1,\sigma}} e^{-\sigma\lambda_1\tau_1 {\rm diag}\para{s^{1,R}_{i,l}}}.\\
&&B^{\sigma}_{l} = e^{-\sigma\lambda_2\tau_2 {\rm diag}\para{s^{2,L}_{i,l}}}   e^{-\tau_2 K^{2,\sigma}} e^{-\sigma\lambda_2\tau_2 {\rm diag}\para{s^{2,R}_{i,l}}}.
\end{eqnarray}
In the above expression, $K^{1,\uparrow}=K^{2,\downarrow}$ denote the hopping matrices associated with the model Hamiltonian (Hubbard model in our example), and $K^{2,\sigma}$ denotes the variational hopping matrix. Here, for simplicity we have assumed explicit BCS pairing is absent (as well as spin conservation), though it is straightforward to take them into consideration and generalize the above relations. The fermion Green's function associated with a fixed Hubbard-Stratonovic binary field configuration $\para{\bf s}$ is:
\begin{eqnarray}
&& G^{\sigma}_f\para{\bf s} = \frac{B^{\sigma}_{N_1}\cdots B^{\sigma}_{N_1/2+1}A^{\sigma}_{N_2}\cdots A^{\sigma}_{1}B^{\sigma}_{N_1/2}\cdots B^{\sigma}_{1} }{\mathbb{1} +B^{\sigma}_{N_1}\cdots B^{\sigma}_{N_1/2+1}.A^{\sigma}_{N_2}\cdots A^{\sigma}_{1}B^{\sigma}_{N_1/2}\cdots B^{\sigma}_{1} }.\cr
&& \label{eq:GF}
\end{eqnarray}
We are now able to use Monte Carlo methods (such as the Metropolis-Hasting algorithm) to sample the Hubbard-Stratonovic fields by considering $\abs{Z_{\up}\para{\bf s}Z_{\dn}\para{\bf s}}$ as the weight of configuration $\bf s$. Moreover, we can employ the Sherman-Morrison-Woodbury identity~\cite{Sherman_1950a,Woodbury_1950a} to efficiently update determinants and temporal Green's function. Also, it is well known that the product of exponential forms in Eq. \ref{eq:Z-1} is numerically unstable and one has to utilize QR decomposition to stabilize them~\cite{Bai_QR_2009a}.
Since each component of the density matrix $\rho$ (i.e., for a fixed Hubbard-Stratonovic configuration ${\bf s}$) is Gaussian and noninteracting, we can use the above relation for the two-point functions to compute four-point and higher order correlation functions by applying Wick's theorem for the corresponding realization of auxiliary fields. By sampling enough important field configurations and averaging over them we can achieve the correlation functions for the interacting problem.

Let us now briefly comment on how to handle $H_s$ terms in Eq. \ref{eq:Trotter-2}. We expect the renormalized coupling constants for the Heisenberg interaction $J_{ij}$ to be small as their bare values are all zero. More precisely, we expect $\tau J_{ij} \ll 1$. Therefore, it is justified to use the conventional Trotter-Suzuki expansion: $e^{-\tau \sum_{ij}J_{ij}S_i.S_j} \approx \prod_{ij}e^{-\tau J_{ij} S_i.S_j} ~+ O\para{\para{\tau J_{ij}}^2}$. Now, recall that $S_i.S_j$ can be decomposed in the hopping channels as: $S_i.S_j = -\frac{1}{2}\hat{\chi}_{ij}^\dag \hat{\chi}_{ij} +\frac{1}{4} n_in_j +\frac{1}{2}n_{i}$, where $\hat{\chi}_{ij} \equiv \sum_{\sigma} c_{i,\sigma}^\dag c_{j,\sigma}$. The second term can be absorbed into the density-density interactions via $V_{ij}\to V_{ij} + J_{ij}/4$ adjustment. Likewise, the third term can be absorbed into the chemical potential term. When $J_{ij} >0$, the resulting decomposition is an attractive interaction in terms of $\hat{\chi}_{ij}$. Hence, we can employ the Hubbard-Stratonovic transformation and rewrite $e^{1/2\tau J_{ij} \hat{\chi}_{ij}^\dag \hat{\chi}_{ij} }$ using an ensemble over continuous or discrete auxiliary fields coupled to fermion hopping operators $\hat{\chi}_{ij}$. For the sake of completeness, we would like to add that $S_{i}.S_{j}$ can also be decomposed in the pairing channel as follows: $S_i.S_j = -\frac{1}{2}\hat{\Delta}_{ij}^\dag \hat{\Delta}_{ij} +\frac{1}{4} n_in_j$, where $\hat{\Delta}_{ij} \equiv c_{i,\uparrow}c_{j,\downarrow}-c_{i,\downarrow}c_{j,\uparrow}$. This decomposition allows an alternative Hubbard-Stratonovic transformation.

\begin{figure*}
\subfigure[]{\includegraphics[height=3.2cm]{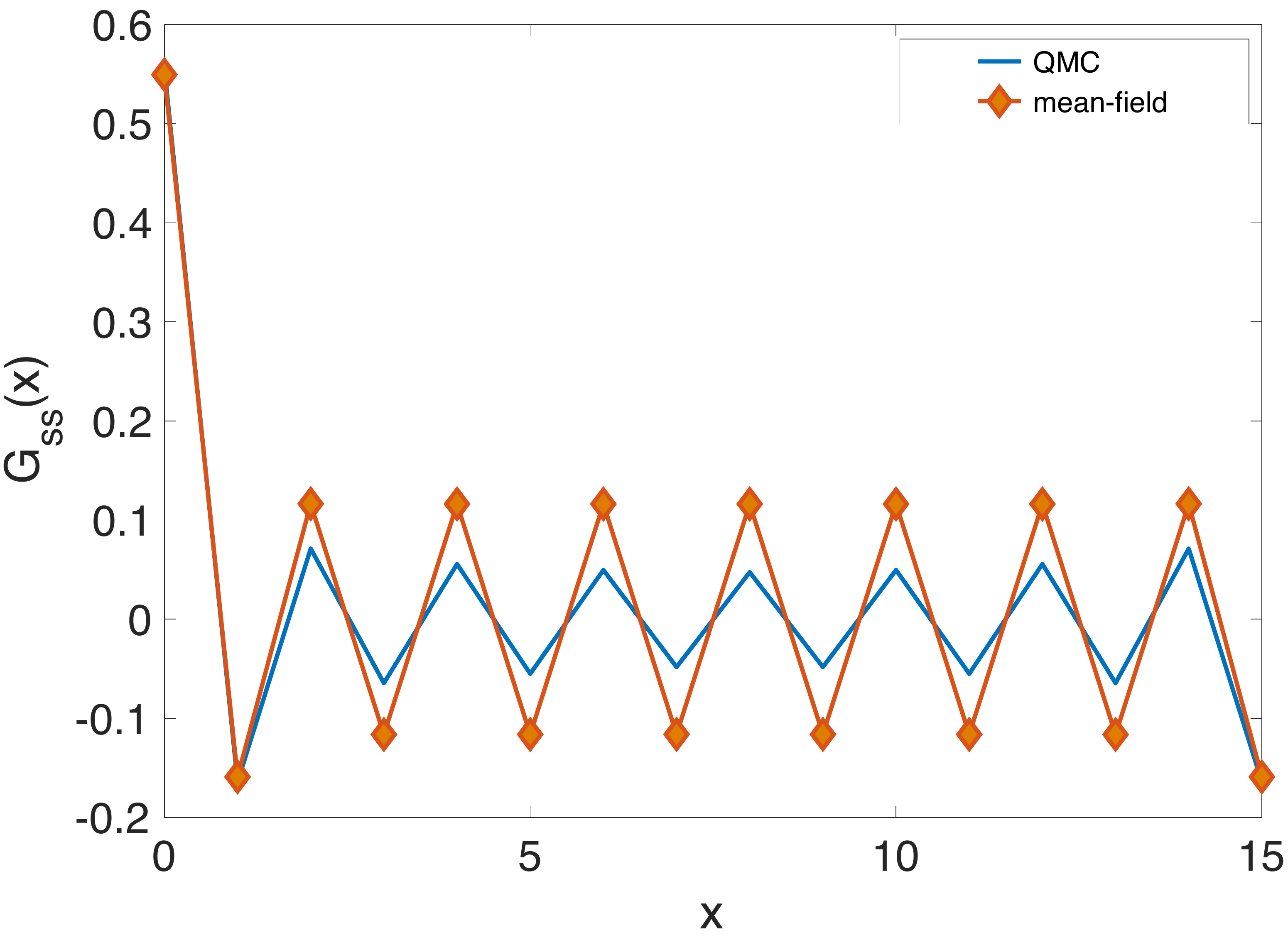}}
\subfigure[]{\includegraphics[height=3.2cm]{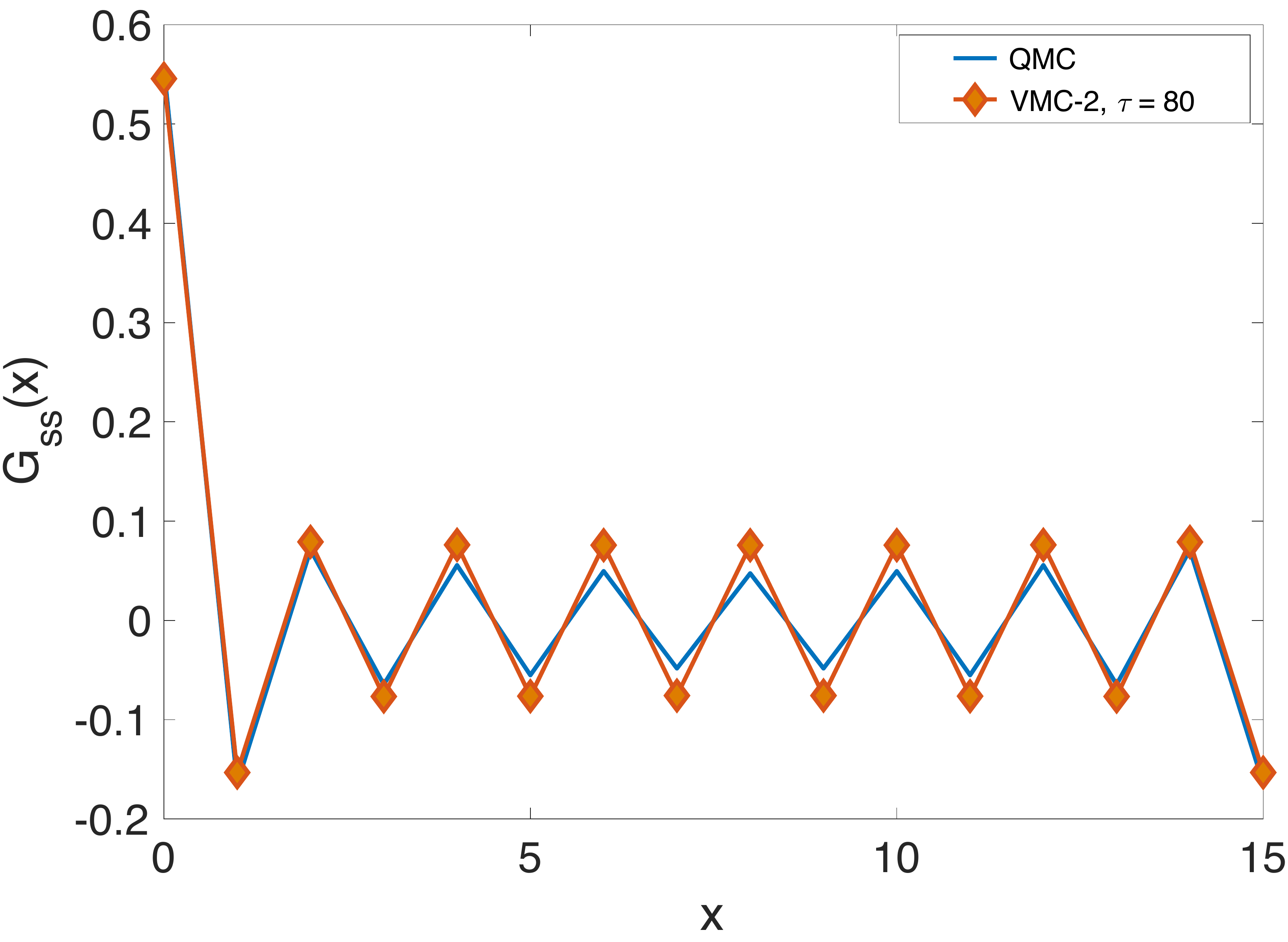}}
\subfigure[]{\includegraphics[height=3.2cm]{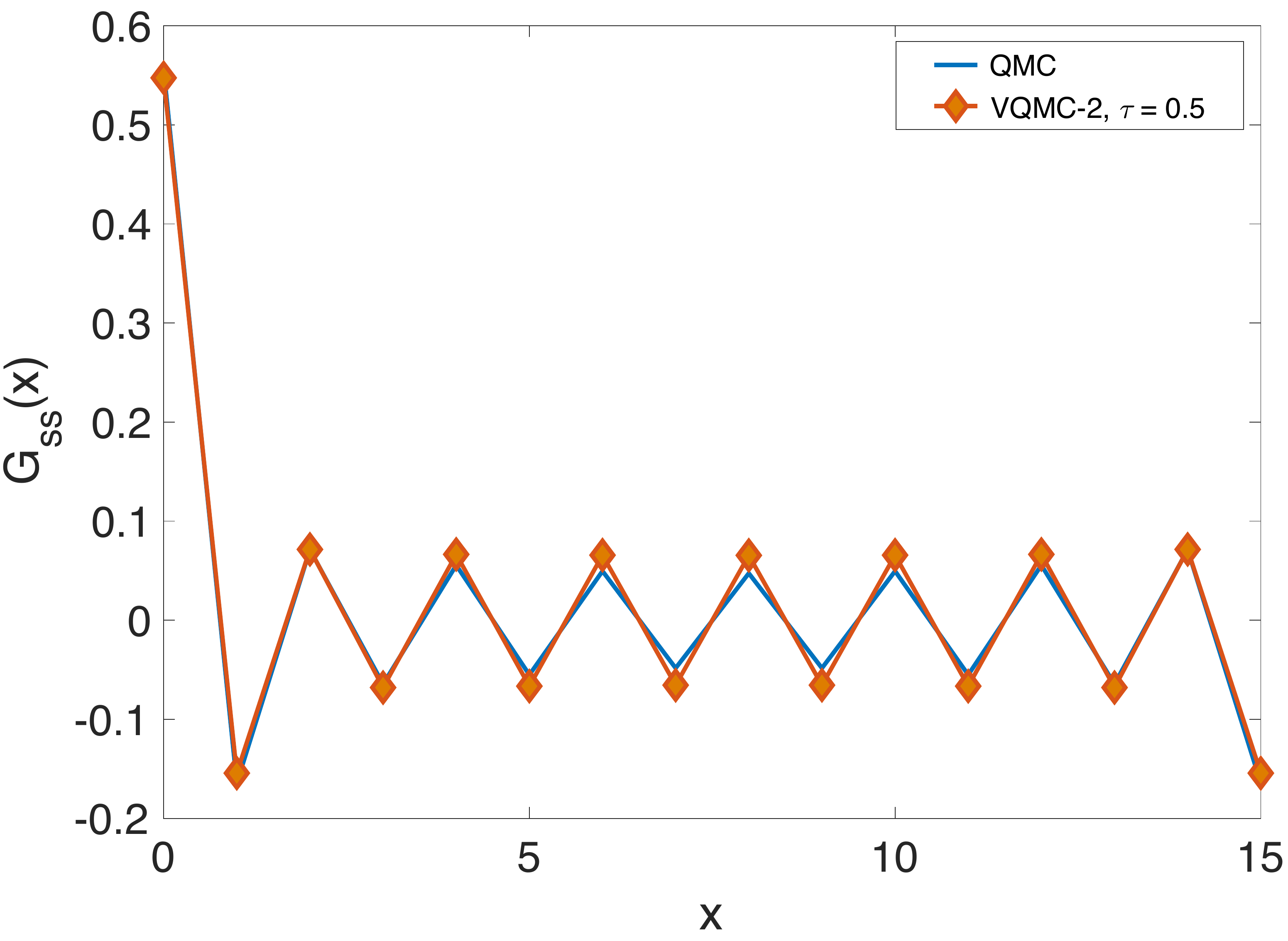}}
\subfigure[]{\includegraphics[height=3.2cm]{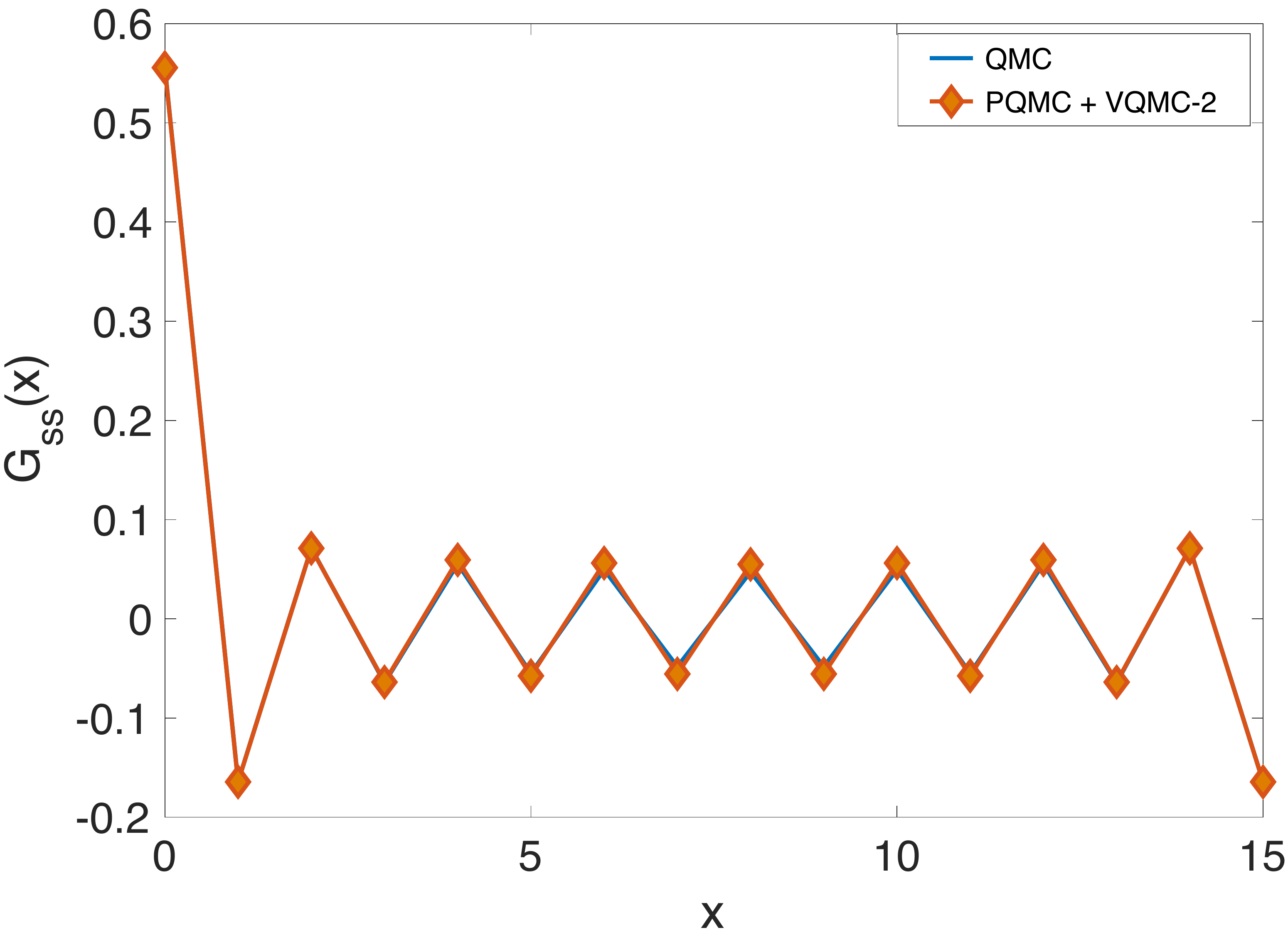}}
\caption{Spin-spin correlation function along $x$ direction obtained by (a) mean-field, (b) VMC-2, (c) VQMC-2, and (d) the hybrid of PQMC and VQMC-2 methods. The solid blue line provides the exact values for reference. Again, we have considered $U=4$ and $16\times4$ dimensions. The hierarchy of accuracies can be seen from these plots. Again, VQMC (at $\beta_2 = 20$) outperforms VMC (at $\beta_2 = 80$) as a result of handling quantum fluctuations better. On the other hand, VQMC can help the PQMC method to achieve satisfactory results even at $\beta_1 = 2$.} \label{fig:Fig_6}
\end{figure*}

\section{IV. Behavior of the sign problem} 
 
In this section, we briefly discuss the origin of fermionic sign problem in DQMC and related methods and argue that our VQMC as well as our modified PQMC combined with VQMC can alleviate the sign problem. The sign problem simply means $Z_{\sigma}\para{\bf s}$ defined in Eq. \ref{eq:Z-1} can take negative values as well depending on the $\bf s$  auxiliary field realization. Since, $Z_{\sigma}\para{\bf s} = {\rm Tr}\para{e^{-c_{i,\sigma}^\dag h^{N}_{ij,\sigma}c_{j,\sigma}}\cdots e^{-c_{i,\sigma}^\dag h^{1}_{ij,\sigma}c_{j,\sigma}}}$ it might seem surprising why the product of individually positive definite terms can give rise to negative values. To understand why that can happen, we must note that there exists $h_{\sigma}^{\rm eff}$ such that
\beq
e^{-c_{i,\sigma}^\dag h^{N}_{ij,\sigma}c_{j,\sigma}}\cdots e^{-c_{i,\sigma}^\dag h^{1}_{ij,\sigma}c_{j,\sigma}} = e^{-c_{i,\sigma}^\dag h^{\rm eff}_{ij,\sigma}c_{j,\sigma}},
\eeq
where $e^{h^{\rm eff}_{\sigma}} = e^{h^{N}_{\sigma}}\cdots e^{h^{1}_{\sigma}}$. However, it is clear that $h^{\rm eff}_{\sigma}$ is not necessarily self-conjugate unless $h^{t}_{\sigma} = \para{h^{N-t}_{\sigma}}^\dag$ which is not true in general as all Hubbard-Stratonovic fields are independent random binary numbers. Therefore, there is no guarantee that $h^{\rm eff}_{\sigma}$ contains real eigenvalues in its SVD decomposition, and thus ${\rm Tr}\para{e^{-c_{\sigma}^\dag h^{\rm eff}_{\sigma}c_{\sigma}}} = \det\para{\mathbb{1}+e^{-h^{\rm eff}_{\sigma}}}$ can have a complex phase. However, for real Hamiltonians the complex phase can be either $1$ or $-1$.  Another way to understand the sign problem is to consider the following imaginary time evolution operator: $U^{\sigma}_t\equiv\prod_{l=1}^{t}e^{-c_{\sigma}^\dag h^{l}_{\sigma}c_{\sigma}}$~\cite{Troyer_SP_2014a}, and its instantaneous many-body lowest right eigenstate denoted as $\ket{\psi\para{t}}$. This many-body eigenstate has $t$ dependence and when $t$ varies slowly, we can define the Berry phase for two consecutive instantaneous eigenstates, namely $e^{i\delta \theta_B(t)}=\left<\psi\para{t+1}|\psi\para{t}\right>$. Since, $h^t_{\sigma}$ depends on the Hubbard-Stratonovic fields which take different values from one imaginary time slice to another, $\ket{\psi\para{t}}$ can have a different phase from its ensuing one. The sign problem occurs when $\left<\psi\para{N}|\psi\para{1}\right>$ (which is correlated with the total Berry phase $\theta_B = \sum_{t}\delta \theta_B(t)$) may take negative values. When the (renormalized) interaction strength $U_{\rm eff}\para{\tau}$ is weak, random parts of $h^t_{\sigma}$ can be neglected, hence the instantaneous eigenstates are close and are all negligible and as a result $\delta \theta_B\para{t}$ are infinitesimal. However, when $N = \beta/\tau \gg 1$, the sum of individual contributions may add up to $\pi$ and cause negative signs. The chance of such events increase exponentially with $\beta$, and that is another reason why the average sign dies off as $e^{-\beta Vf}$ with $\beta$. However, there are two other effects that can ameliorate the sign problem. For instance, as we discussed previously and Figs. \ref{fig:Fig_4} and \ref{fig:Fig_7} suggest, $U_{\rm eff} \ll 1$ for $\tau \sim O(1)$ and beyond. Additionally, when symmetry breaking terms such as staggered magnetization are allowed, the random onsite Hubbard-Stronovic fields are masked by the mean-field terms and their chance to negate the partition functions is exponentially suppressed. Moreover, as Fig. \ref{fig:Fig_2_3} shows, it is energetically favorable to consider such symmetry breaking terms for finite imaginary time steps though their strength is attenuated upon decreasing $\tau$.

To summarize, there are two reasons to push down the onset temperature for the emergence of the sign problem in our VQMC and PQMC methods. Firstly, considering imaginary time steps larger than that of the DQMC method surpasses the renormalized onsite Hubbard coupling $U_{\rm eff}$ as $\tau^{-1}$. Secondly, for finite imaginary time steps, we find a non-vanishing optimal value for the staggered magnetization at half filling and other symmetry breaking terms in general. These two observations hand in hand ameliorate the sign problem and keep the average sign high. As a result, in PQMC, the VQMC trial density matrix part of the method is nearly sign free and the sign problem is entirely due to the projection part namely $e^{-\beta_1 H/2}$ factor. Thus, despite its remarkable performance and accuracy, the sign problem behaves as $e^{-\beta_1 V f}$ instead of $e^{-\para{\beta_1+\beta_2}Vf}$. Finally, since the VQMC ansatz was already close to the exact ground-state, the minimum required projection time $\beta_1/2$ can be surprisingly short (see Fig. \ref{fig:Fig_5}).

\section{V. Results} In this section we first benchmark our unifying VQMC algorithm, which encompasses VMC and DQMC as its two extreme limits (upon varying imaginary time steps $\tau$), and later employ it to feed PQMC and present the corresponding results. In this paper, we focus on the half-filled Hubbard model on the square lattice with $\para{N_x,N_y} = \para{16,4}$ linear dimensions. We consider $U \in \left[1,6\right]$ range for the onsite Hubbard interaction. For these model Hamiltonians, the finite temperature DQMC method is sign-free and can provide exact results up to any desired accuracy by considering large $\beta$ and sampling over enough Hubbard-Stratonovic field configurations. In our simulations, we have assumed the following imaginary time steps:  $\tau = \para{1/20, 1/2,1,2,4,8,\beta}$, and considered $\beta \leq 80$ (in units of $t_1^{-1} = 1$). These time steps correspond to (i) VMC for $\tau = \beta$, (ii) DQMC for $ \tau = 1/20$ (iii) VQMC for $ \tau = \para{1/2,1,2,4,8}$ values, respectively. To gain a better insight on how accurate our VMC and VQMC methods are, we also present the mean-field estimations for various quantities besides exact results obtained from DQMC. We finally present our results for the PQMC method with a trial density matrix obtained through the VQMC method.

\begin{figure}
\includegraphics[height=3.1cm]{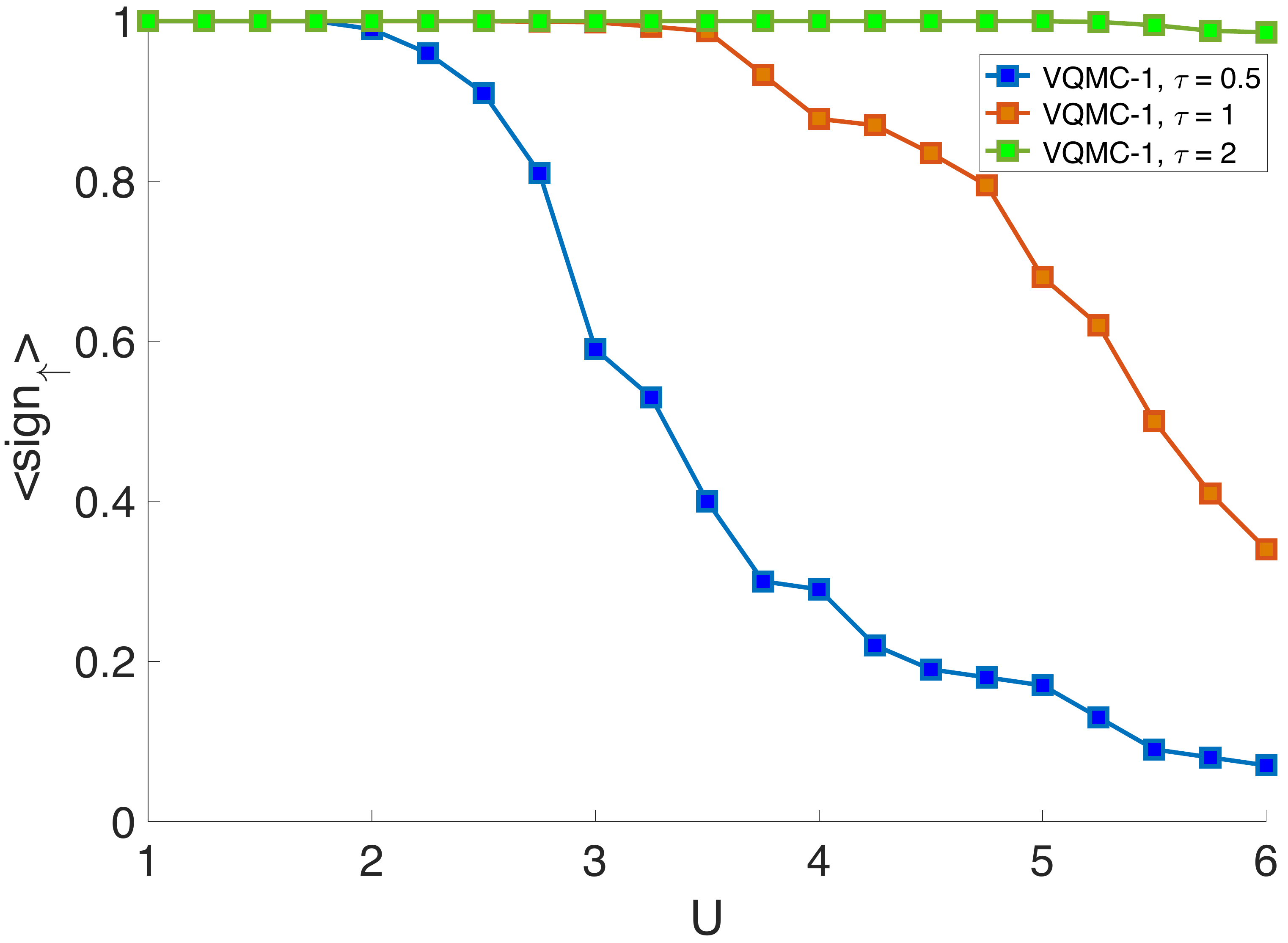}~~~
\includegraphics[height=3.1cm]{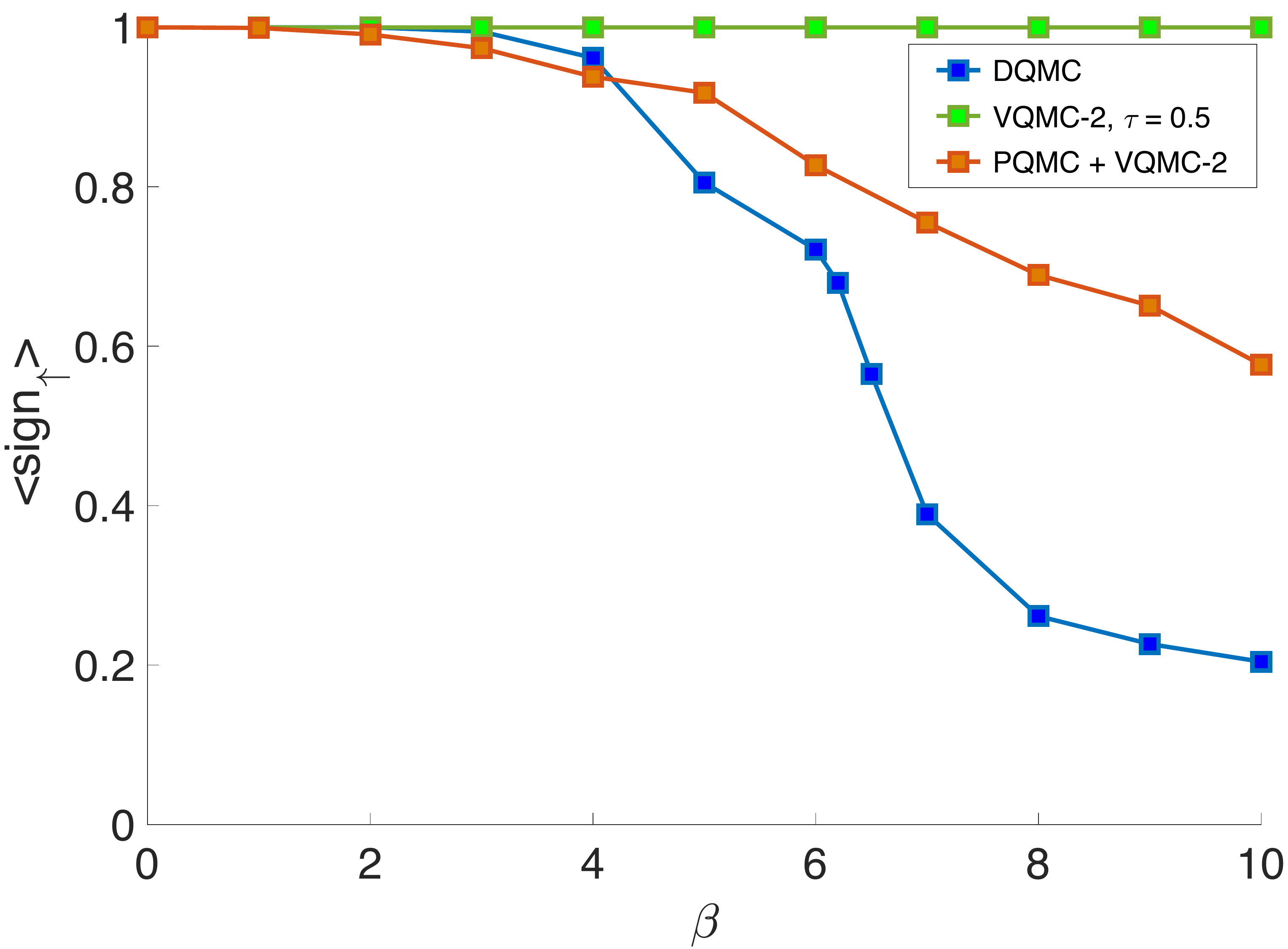}~~~
\caption{Average sign of spin up determinants versus $U$ and inverse temperature $\beta$. This figure provides useful insight for the behavior of average overall sign away from half-filling. (a) Average sign versus $U$ at $\beta_2 = 20$ for different choices for imaginary time step $\tau$. Here, we have plotted our results for VQMC with one tuning parameter. The average sign does not fluctuate at $\beta_2 = 20$ for VQMC when we allow two variational parameters, i.e., staggered magnetization besides $U_{\rm eff}$. (b) Average sign versus $\beta$ at $U=4$. This plot shows that the average sign is most severe for DQMC. In PQMC (with a VQMC-2 input) we can already probe ground-state properties around $\beta_1 = 2$  (see Fig. \ref{fig:Fig_4} ) at which the average sign is well behaved. The average sign for VQMC-2 is constant within the considered temperature range. } \label{fig:Fig_7}
\end{figure}

For simplicity, in this work we only consider up to two variational parameters. More explicitly, we have considered $J_{r} = 0$, $V_{r\neq 0} = 0$, and kept $V_{0}$ which denotes the renormalized onsite Hubbard interaction. Moreover, motivated by the mean-field approximation we consider the following form of the kinetic term which contains one variational parameter $m$ associated with the staggered magnetization:
\beq
H_{M} = -t_1\sum_{\braket{ij},\sigma} c_{i,\sigma}^\dag c_{j,\sigma} -m \sum_{i,\sigma} \para{-1}^{i_x+i_y}\sigma n_{i,\sigma}.\label{eq:MF-1}
\eeq
 In fact, one can go beyond the above simple sparse form of the variational parameters and improve the accuracy of our results by considering additional variational terms e.g., further neighbor hoppings, nearest neighbor Heisenberg and density-density interactions.  
 
In the following computations, we have considered enough spacetime sweeps over the Hubbard-Stratonovic fields to ensure the statistical error in energy estimation remains below $0.005$. In our plots, VMC-1 (VQMC) denotes our VMC (VQMC) ansatz with one tuning parameter, and VMC-2 (VQMC-2) indicates those with two variational parameters. 

In Fig. \ref{fig:Fig_1}(a) we present the estimated ground-state energies obtained via the standard finite temperature DQMC method for $\beta = 20$. We consider them as exact values for the ground-state energy through verifying their stability against decreasing temperature (increasing $\beta$) further. Furthermore, to have an idea how bad the sign problem can be, we report the average sign of spin up determinants within DQMC at $\beta = 20$ in Fig. \ref{fig:Fig_1}(b).  Note that the average sign diminishes as we increase $U$.

In Fig. \ref{fig:Fig_2_3}(a) we compare the estimated ground-state energy (extrapolated to $\beta = \infty$) relative to the ground truth for a range of $U$ values obtained through applying various methods. In this figure we have set $m=0$, and only $U_{\rm eff}$ can vary. In Fig. \ref{fig:Fig_2_3}(b) we plot the optimal values of $U_{\rm eff}$ for these models given $m=0$ constraint. Similarly, Figs. \ref{fig:Fig_2_3}(c) and \ref{fig:Fig_2_3}(d) present estimations for the ground-state energy (relative to the exact one) and the corresponding renormalized onsite Hubbard interaction by optimizing both variational parameters namely $U_{\rm eff}$ and $m$. 
These two figures suggest a considerable improvement to the mean-field approximation by considering VMC. Furthermore, VQMC improves the results obtained from VMC further which was achieved by taking quantum fluctuations into consideration.

In Fig. \ref{fig:Fig_4}, we plot the renormalized onsite Hubbard interactions as a function of $\tau$ for $U=4$. As we expect, $U_{\rm eff} = U$ for $\tau = 1/20$ (i.e., in DQMC), while it decays as $1/\tau$ for $\tau \geq 1/2$. 

In Fig. \ref{fig:Fig_5}, we compare the estimated ground-state energy of the PQMC method through considering various choices for the trial state/density matrix. This result suggests the VQMC as the best choice for the trial density matrix to feed in PQMC, since a projection time ($\beta_1$) as short as 1 can already result in a highly accurate estimation for the ground-state energy. For the mean-field trial state on the other hand, we must consider $\beta_1 = 3$ to reach that accuracy. We would like to stress that all these results are obtained by allowing at most two variational parameters in VQMC. Using more variational parameters can reduce the threshold of $\beta_1$ ($\beta^{ th}\para{\epsilon,\rm PQMC}$) further.

Fig. \ref{fig:Fig_6}, compares the spin-spin correlation functions for $U=4$ obtained via different approaches. Again, we see that VQMC yields satisfactory results by comparing it with those of the DQMC. Also, PQMC achieves a highly accurate result even though we have considered $\beta_1=1$. 

Finally, Fig. \ref{fig:Fig_7} presents the average sign of spin up fermion determinants for $U=4$ versus temperature as well as average energy of the system. These plots show how the average sign (for spin up fermions) can increase exponentially using our modified PQMC (fed by VQMC as its trial density matrix) despite its remarkably high accuracy. This result suggests that our modified PQMC can uncover the ground-state properties of the doped Hubbard model by allowing to access and probe lower temperatures.

\section{VI. Symmetry breaking, competing orders,  and ergodicity} 
The quadratic Hamiltonian considered in Eqs. \ref{eq:Trotter-1} and \ref{eq:MF-1} can host terms that break various symmetries associated with the Hubbard model. For example, the staggered magnetization term breaks the spin SU(2) symmetry down to its $Z_2$ subgroup. It also breaks the lattice translation as well as $C_4$ rotational symmetries and enlarges the unit cell accordingly. However, we would like to stress that considering such terms are not necessary in general. For instance, considering a nonzero value for the $J_1$ term in Eq. \ref{eq:Trotter-2}, we can compensate the effects of staggered magnetization considered in Eq. \ref{eq:MF-1}. It can also take quantum fluctuations (such as Goldstone modes) around the symmetry breaking terms into consideration for free. This however, depends on how stable that symmetry breaking phase is compared to other potentially competing orders. If the estimated ground-state energies of those competing symmetry breaking phases are well separated, we may simply consider a symmetry breaking quadratic form. Otherwise, it will be more reasonable to keep the quartic Heisenberg terms to allow for more complicated spin phases, e.g., spin liquids.

\section{VII. Summary and Discussion} 
In this work we presented a unified framework for several QMC approaches. We demonstrated that they all can be understood using our generalized Trotter-Suzuki decomposition (see Eq. \ref{eq:Trotter-2}). Based on this understanding we developed a novel technique dubbed as VQMC which paves the way between VMC and DQMC and interpolates between them by smoothly varying imaginary time steps from zero to infinity. We showed that this novel method is more accurate than VMC, captures important (low energy) quantum fluctuations, and can give access to low temperature due to its better behavior for the sign problem.  We showed that our VQMC can serve as the best available trial state for the PQMC upon which we can achieve ground-state properties even after short projection time. We investigated various aspects of these related techniques.

There are still several important steps to be taken in future. In this paper, we focused on the unfrustrated Hubbard model at half filling to benchmark our algorithm since we can find the exact solutions using the conventional DQMC algorithm. However, such proposals are more needed away from half filling or for frustrated Hamiltonians that suffer from fermionic sign problem. Hence, it is interesting to see what we can learn from this new approach when applied to doped Hubbard model whose reliable solution (even approximate) is still absent. 

In this work, we considered two variational parameters at most, and achieved satisfactory results. However, the VQMC itself can be significantly improved by allowing more variational terms in Eq. \ref{eq:Trotter-2} such as the longer range Jastrow density-density or Heisenberg spin-spin interactions. We have already seen that considering two variational parameters gives rise to much more accurate results than a single variational parameter (see Fig, \ref{fig:Fig_2_3} for example).

It is worth noting that our algorithm can be useful for the finite temperature DQMC as well. It obviates the need to extrapolate to $\tau \to 0$ to kill the Trotter errors. Instead, we need to consider short enough (imaginary) time steps (e.g., $\tau = 1/4$ or $\tau = 1/8$) and optimize $U_{\rm eff}$ which will turn out to be slightly less than $U$. Fig. \ref{fig:Fig_4} shows that even for $\tau=1/4$, $U_{\rm eff} \neq U$. For example, we observed that for $U=4$ and $\tau = 0.25$, considering $U_{\rm eff} \approx 3.5$ yields results closer to those of $\tau = 0.05$ than $U_{\rm eff} = 4$. 

Finally, we would like to mention that our two-time representation in Eq. \ref{eq:rho-PQMC-1} can be straightforwardly generalized to more general situations such as $e^{-\beta_1 H/2} e^{-\beta_2 H/2}e^{-\beta_3 H}e^{-\beta_2 H/2}e^{-\beta_1 H/2}$, where $\beta_1 \ll \beta_2 \ll \beta_3$ and then consider generalized Trotter-Suzuki decomposition for $\tau_1 \ll 1$, $\tau_2 \sim O(1)$, and $\tau_3 \gg 1$ imaginary time steps. Such additional decorations can enhance the computational time, accuracy, and improve the average sign further.

\section*{Acknowledgements}
We thank Edwin Huang, Sai Iyer, Zohar Nussinov, Brian Moritz, Christian Mendl, and Yoni Schattner for useful discussions. A.V. acknowledges the Gordon and Betty Moore Foundation's EPiQS Initiative through Grant GBMF4302.

%

\end{document}